\documentclass[%
 a4paper,
 reprint,  
 unsortedaddress,
 amsmath,amssymb,
 prb,
]{revtex4-1}

\usepackage{graphicx}
\usepackage{color}
\usepackage{dcolumn}
\usepackage{bm}
\usepackage{dcolumn}
\usepackage{hyperref}

\usepackage{epstopdf}
\begin{document}

\title{Comparative study of structural and electronic properties
of GaSe and InSe polytypes}
\author{Juliana Srour}
\author{Micha\"el Badawi}
\altaffiliation{Present address: Laboratoire Physique et Chimie Th\'eoriques 
(LPCT, UMR CNRS UL 7019), Universit\'e de Lorraine, Rue Victor Demange,
F-57500 Saint-Avold, France}
\affiliation{LCP-A2MC, Institute Jean Barriol, Universit{\'e} de Lorraine, 1 Bd Arago, F-57078 Metz, France}
\author{Fouad El Haj Hassan}
\affiliation{Universit\'e Libanaise -- Facult\'e de Sciences (I),
LPE -- Laboratoire de Physique et d'Electronique,
Campus Rafic Hariri -- Hadath, Beirut, Lebanon}
\author{Andrei Postnikov}
\email[Corresponding author: ]{andrei.postnikov@univ-lorraine.fr}
\affiliation{LCP-A2MC, Institute Jean Barriol, Universit{\'e} de Lorraine, 1 Bd Arago, F-57078 Metz, France}
\date{\today}
\begin{abstract}
Equilibrium crystal structures, electron band dispersions
and band gap values of layered GaSe and InSe semiconductors,
each being represented by four polytypes, are studied via first-principles
calculations within the density functional theory (DFT). A number of practical algorithms
to take into account dispersion interactions are tested, from
empirical Grimme corrections to many-body dispersion schemes. 
Due to the utmost technical accuracy achieved in the calculations, 
nearly degenerate energy-volume curves of different polytypes are resolved,
and the conclusions concerning the relative stability of competing polytypes
drawn. The predictions are done as for how the equilibrium between
different polytypes will be shifted under the effect of hydrostatic pressure.
The band structures are inspected under the angle of identifying features
specific for different polytypes, and with respect to modifications of
the band dispersions brought about by the use of modified Becke-Johnson (mBJ)
scheme for the exchange-correlation (XC) potential. As another way to 
improve the predictions of band gaps values, hybrid functional calculations
according to the HSE06 scheme are performed for the band structures,
and the relation with the mBJ results discussed. 
Both methods nicely agree with experimental results and with 
state-of-the-art GW calculations. Some discrepancies are identified in cases of
close competition between the direct and indirect gap (e.g., in GaSe);
moreover, the accurate placement of bands revealing relatively localized states
is slightly different according to mBJ and HSE06 schemes. 
\end{abstract}
\maketitle

\section{Introduction}
\label{sec:intro}
Layered nature of III-VI semiconductors, known since long,
experiences in the last decades a renaissance of interest, related to
two-dimensionality of properties and promising applications.
The structure of these materials is such that cations 
(Ga or In, the the present work) are bonded, in the tetrahedral coordination,
to another similar cation and to three anions (Se, in the present work).
The anions place themselves in hexagonal arrangement at two surfaces
of what is in the following referred to as \emph{double layer}, each anion
being bonded to three cation beneath (within the layer). The inner cation-cation bond
directs at normal to the surface, and the anions at the two opposite surfaces
are (in all known structure modifications) in the eclipsed (wurtzite-like) 
configuration.\footnote{For isolated In-chalcogenide double layers\cite{PRB89-205416}
and for Ga/In-chalcogenide bi(double)layers,\cite{JChemPhys147-114701}
a staggered configuration was equally probed in theory calculations.}
The double layers can be stockpiled in a variety of sequences within the globally
hexagonal symmetry. The anions' valences being saturated, there is formally
no covalent bonding between the adjacent double layers, hence the role of 
dispersion interactions (DI) in holding the layer system together is large.

From the point of view of experiments or applications, the step of bulk materials'
characterization by spectroscopies and other techniques being now history, 
the modern interest for these materials is largely fed by possibilities 
to exfoliate\cite{ACSnano8-1263,2DMater4-025043} or 
grow\cite{SciRep4-5497,NanoLett13-2777,TSF542-119} single layers, 
dope them\cite{PCCP17-10737} or otherwise distort, and bring together with other layered materials
into fancy heterostructures or devices.\cite{AdvMat24-3549,JACS131-15602,JChemPhys147-114701} 

From the point of view of first-principles studies,
history are calculations of electronic structure of a single layer,
or of singular polytypes, addressing particular problems or particular experiments.
Nowadays as practical schemes to include the DI on top of, or within,
the calculations done with the density functional theory (DFT)
do gradually become routine, it seems interesting and important to revise 
the accuracy with which these methods would address the structures and
relative stabilities of (presumably quite competitive) polytypes.
Another point of practical interest which can be addressed by calculations
is the estimation of fine variations of band gaps over structurally close materials,
that is a clue for tuning the optical properties by structure engineering in
the desirable direction. Systematically underestimated, due to well-known 
``deficiency'' of the exchange-correlation (XC) potential in ``traditional'' 
DFT calculations using, say, the local density approximation (LDA)
or generalized gradient approximation (GGA),
yet available with high accuracy from much more demanding GW calculations, 
the band gaps seem to be nowadays fairly well reproducible 
within schemes which require only moderate intervention into
the DFT calculation routine. Such schemes include namely the hybrid functionals
(which admix exact exchange into a DFT XC functional)
and ``meta-GGA'' techniques (which express the XC potential in terms of 
further parameters than the charge density and its gradient). 

In the present work, we offer a comparative analysis of first-principles
predictions concerning GaSe and InSe semiconductors along the four axes of comparison:
$(i)$ critical assessment of different schemes to include the DI into
the calculations, in view of obtaining utterly accurate description
of the crystallographic parameters;
$(ii)$ comparison of two currently used schemes, -- a realization of meta-GGA
known as ``modified Becke-Johnson'' (mBJ) formula for the XC 
potential\cite{PRL102-226401} and the Heyd-Scuseria-Ernzerhof (HSE)
hybrid functional,\cite{JChemPhys118-8207} --
to obtain band structures and band gaps in good agreement with experiment;
$(iii)$ comparison of GaSe and InSe as systems structurally and chemically close
yet differing in their degree of covalence;
$(iv)$ for each system -- a comparison throughout four polytypes, characterized
by delicate differences in their crystal structures and extremely close in
energy / stability preferences. A resolution of existing differences to reveal
reliable trends demanded an utter care in technical precision of calculations.

Moreover, as yet another line of comparison, two different calculation methods,
WIEN2k\cite{wien2k} and VASP,\cite{vasp}
have been used and, in fact, tested against each other on, in part, similar tasks;
this however did not lead to any general conclusion in favor of one or
the other. The satisfactory agreement of results, provided the technical
prerequisites for sufficiently high accuracy are employed within the method in question,
gave us the necessary confidence in the trends discussed.

The work is organized as follows. Section~\ref{sec:struc} explains the structures
of polytypes, section~\ref{sec:previ} sets the context of earlier calculations
and important experiments,
section~\ref{sec:method} specifies the methods within the DFT which are of
special interest for the present study. Further on, the new results come
arranged by topics, with corresponding discussion:
Section~\ref{sec:optim} deals with optimized crystallographic parameters
and performance of different schemes to include the DI;
section~\ref{sec:eq_state} discusses the relative stabilities of polytypes
in the context of energy/volume curves;
section~\ref{sec:band_gaps} addresses band structures of polytypes
obtained with two different approaches and compares the resulting band gaps 
with experimental data. Section~\ref{sec:conclu} concludes the discussion.

\begin{figure}[b] 
\centerline{\includegraphics[width=0.96\linewidth]{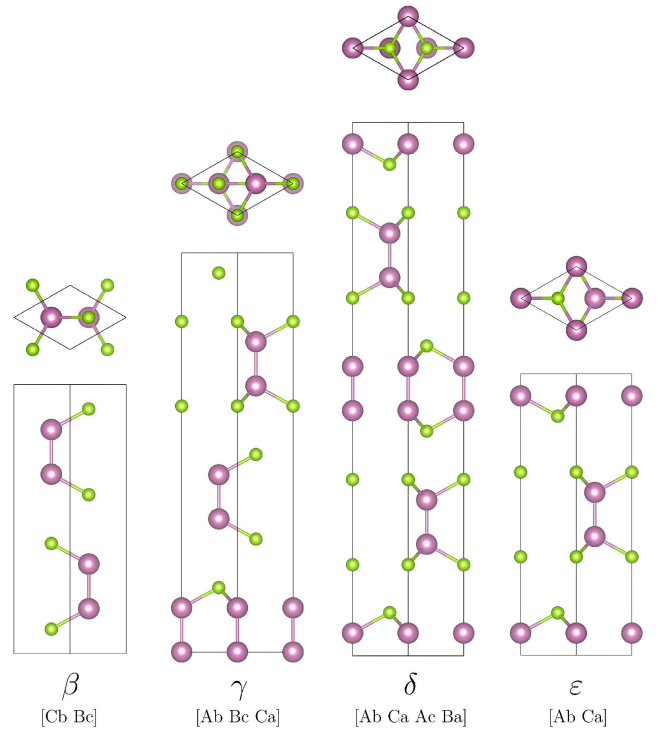}}
\caption{\label{fig:InSe_struc}
Side and top view of the unit cells of $\beta$, $\gamma$, $\delta$ and $\varepsilon$
polytypes of GaSe or InSe. Large circles: cations, small circles: anions. 
The stacking notation at the bottom is explained in the text.}
\end{figure}  

\begin{table*}[th!] 
\caption{\label{tab:polytyp}
Crystal structure definitions for four polytypes of III-VI binaries.
The $z$ coordinates are indicated
as for GaSe, after Kuhn \emph{et al.}\cite{PSSA31-469}}
\smallskip
\begin{ruledtabular}
\begin{tabular}{cccc@{\hspace*{18pt}}c}
\parbox[c]{3.0cm}{Polytype;\\*[-2pt]stacking order} & 
\rule[-10pt]{0mm}{22pt}Space group & \parbox[c]{3.0cm}{Wyckoff\\*[-1pt] positions} & 
$z_{\rm (Ga)}$ & $z_{\rm (Se)}$ \\ 
\hline 
\rule[-6pt]{0mm}{16pt}
$\beta$ [Cb$\,$Bc] & $P6_3/mmc$ (194) & 4(f) & $\frac{1}{4}\!-\!0.07$ & $\frac{3}{4}\!-\!0.16$ \\
\hline 
\rule[-2pt]{0mm}{12pt}
     & & 2(g) & 0.075 & $\frac{1}{2}\!+\!0.15$  \\*[-0pt]
$\varepsilon$ [Ab$\,$Ca]& $P\bar{6}m2$ (187) & 2(h) & $\frac{1}{2}\!+\!0.07$ \\*[-0pt]
\rule[-4pt]{0mm}{12pt}
 & & 2(i) & & 0.15  \\
\hline 
\rule[-5pt]{0mm}{15pt}
$\gamma$ [Ab$\,$Bc$\,$Ca] & $R3m$ (160) & 
3(a) & $-$0.05; $+$0.05 & $\frac{2}{3}\!-\!0.1$; $\frac{2}{3}\!+\!0.1$ \\
\hline 
\smash{\raisebox{-.5\normalbaselineskip}{$\delta$ [Ab$\,$Ca$\,$Ac$\,$Ba]}} &
\smash{\raisebox{-.5\normalbaselineskip}{$P6_3mc$ (186)}} 
\rule[0pt]{0mm}{10pt}
   & 2(a) & $-0.038$; $+0.039$ & 
            $\frac{1}{4}\!-\!0.071$; $\frac{1}{4}\!+\!0.078$ \\*[2pt]
 \rule[-4pt]{0mm}{10pt}           
 & & 2(b) & $\frac{1}{4}\!-\!0.038$; $\frac{1}{4}\!+\!0.038$ &
            $\frac{1}{2}\!-\!0.075$; $\frac{1}{2}\!+\!0.075$ \\
\end{tabular}
\end{ruledtabular}
\end{table*} 

\section{Crystal structures of different polytypes}
\label{sec:struc}
Crystal structures of the four polytypes addressed in the present work
have been systematized are refined (for the GaSe compound) by Kuhn \emph{et al.}\cite{PSSA31-469};
Likforman and Guittard \cite{ComRenAcadSciC279-33} reported the lattice parameters
of $\gamma$-InSe. The latter phase being in fact rhombohedral, it is shown,
among the other polytypes, in the hexagonal setting in Fig.~\ref{fig:InSe_struc}.
All polytypes have unit cells which are similar in projection onto the hexagonal plane,
but differ in the number and lateral placement of stacked double layers. A compact notation
to distinguish the polytypes would be to use a/b/c letters to mark three different
sites in projecting the atoms onto two-dimensional hexagonal lattice,
reserving the uppercase letters for cations and lowercase for anions. Since anions are
always in the eclipsing configuration, a two-letter code suffices to pinpoint 
a double layer, thus extending an (ambiguous) single-letter stacking-sequence
labeling used in Ref.~\onlinecite{PSSA31-469}. The repeated sequence of double layers,
included in square brackets, is indicated in Fig.~\ref{fig:InSe_struc} for each polytype.
In principle, an infinite number of stacking sequences can be constructed,
respecting the simple rule that an anion should never be in the same site
with its closest cation, nor with the adjacent-layer anion.   

Space groups and Wyckoff positions with representative $z$ coordinates (corresponding
to GaSe after Ref.~\onlinecite{PSSA31-469}) are given in Table~\ref{tab:polytyp}. 
The $z$ coordinates in this table are formatted here so as to emphasize the splitting of cation
or anion positions from the median planes of the corresponding double layer.
We note that this splitting amounts, in all structures, to approximately ${\pm}0.15$
for cations and ${\pm}0.3$ for anions, in terms of the $c$ parameter per
double layer. This reveals a relative robustness of the double layer,
the basic structure element differently stacked in different polytypes.
There is a misprint in the original Table of Kuhn \emph{et al.}\cite{PSSA31-469}
for the $\varepsilon$ phase, an anion being attributed to (2h) instead of (2i) position;
this is corrected in Table~\ref{tab:polytyp} (consistently with Fig.~1 of 
Kuhn \emph{et al.}\cite{PSSA31-469} and with the most of structure descriptions since then).

\section{Previous studies}
\label{sec:previ}
Experimental works in angle-resolved photoemission by 
Larsen \emph{et al.}\cite{PRB15-3200} and Amokrane \emph{et al.}\cite{JPCM11-4303} on InSe,
Thiry \emph{et al.}\cite{SSC22-685} and Plucinski \emph{et al.}\cite{PRB68-125304} on GaSe,
as well as angle-resolved inverse photoemission by Sporken \emph{et al.}\cite{PRB49-11093}
on both GaSe and InSe provided precious benchmarks for many subsequent calculations
of electron band dispersions.

After the initial wave of parameter-dependent (using tight-binding models, or semiempirical)
calculations on different phases of GaSe and InSe since the end of 1970s,\cite{PRB13-3534,%
PRB14-424,JPhysC10-1211,SSC27-1449,NuovoCimB51-154,JPhysC12-1625,JPhysC12-4777,PhysicaBC105-324}
the III-VI semiconductors regained interest since about mid-1990s for \emph{ab initio}
studies within the DFT.\cite{PRB48-14135,PRB57-3726,JPCM11-6715,ChinPhysLett23-1876,PRB84-085314}
Of the works done within several last years, Ghalouci 
\emph{et al.}\cite{CompMatSci67-73,CompMatSci124-62}
calculated the equations of state of $\beta$- and $\varepsilon$- GaSe\cite{CompMatSci67-73}
and InSe\cite{CompMatSci124-62} in comparison with other phases (typical for ``conventional''
semiconductors but too high-energetic for the III-VI systems),
using the WIEN2k method in combination with the GGA.
Ma \emph{et al.}\cite{PCCP15-7098} provided band structure calculations for GaSe
(along with GaS, using the VASP code and GGA) as bulk crystal (lattice parameters optimized)
and as a multilayer system (with 1 to 4 double layers). 
Olgu{\'{\i}}n \emph{et al.}\cite{EurPhysJB86-350} relaxed the structure of $\gamma$-InSe
and $\varepsilon$-GaSe, using the WIEN2k and GGA (applying some additional efforts
in the study of band gaps). 
Rak \emph{et al.}\cite{JPhChSol70-344} simulated, via WIEN2k and VASP calculations,
the electronic structure of pure $\beta$-GaSe and (within the supercell approach)
that containing point defects.
Zhang \emph{et al.}\cite{PhysicaB436-188} calculated equilibrium structure, 
elastic and optical properties of $\varepsilon$-GaSe by a 
pseudopotential planewave (PP-PW) method (CASTEP code).
Rybkovskiy \emph{et al.}\cite{PRB90-235302} calculated band structures
of $\beta$- GaSe, InSe, and GaS in dependence on number of double layers,
using another realization of the PP-PW formalism (Quantum Espresso)
and several flavors of GGA for comparison, with spin-orbit taken into account. 
Debbichi \emph{et al.}\cite{JPhysChemLett6-3098} optimized crystal structure
of $\beta$- and $\gamma$-InSe using the VASP code (with GGA and the Grimme's
correction to simulate DI, see Sec.~\ref{sec:method}),
inspected the effect of inclusion of the spin-orbit interaction on the electron bands,
and calculated the band gaps by the GW method.

A number of recent theory works primarily addressed the optical, elastic, or vibrational
properties of GaSe and InSe under hydrostatic pressure, or under stress,
often in the context of comparison with experimental 
studies.\cite{PRB63-125330,PRB66-085210,PRB71-125206,PSSB244-244,PRB77-045208,%
PRB70-125201,Semicon44-1158,PRL108-266805,JPhChSol74-1240,PCCP15-7098,%
PhysSolState46-179,CompMatSci124-62} 
Even as simulations under pressure are not by themselves our interest in the present study,
the cited works may provide useful references concerning the equation of state curves 
(energy vs volume) over a broad range around equilibrium, as well as the numerical results
at equilibrium.
Adler \emph{et al.}\cite{PRB57-3726} reported elastic constants and phonon dispersion
in $\varepsilon$-GaSe.

The issue of band gap and its assessment in GW calculations has been addressed
by Ferlat \emph{et al.}\cite{PRB66-085210} for $\gamma$-InSe,
by Debbichi \emph{et al.}\cite{JPhysChemLett6-3098} for $\beta$ and $\gamma$-InSe,
by Rybkovskiy \emph{et al.}\cite{PRB84-085314} for GaSe,
by Ayadi \emph{et al.}\cite{JChemPhys147-114701} for Ga- and In-chalcopyrite bilayers. 
Olgu{\'{\i}}n \emph{et al.}\cite{EurPhysJB86-350} 
discussed the band structures calculated for $\gamma$-InSe and $\varepsilon$-GaSe 
(using the WIEN2k code with GGA and mBJ) in the context of available GW calculations.
Wei An \emph{et al.}\cite{JChemPhys141-084701} discussed the band gap, as obtained
by different methods for $\varepsilon$-GaSe (and also in $\beta$-GaS), offering an overview
across other available results and implementations. Notably an excellent agreement
has been found between the GW and the mBJ predictions for the band gap and for
the band structure in the latter's vicinity.

Despite so many calculations done by state-of-art methods, one can note
the following insufficiencies that justify the necessity of our present study:
($i$) The works usually address the InSe or GaSe compound
in one particular phase, or, at most, comparing two structures. Considerable efforts
were spent on comparison with ``irrelevant'' structures (typical for other 
semiconductors but not for III-VI). 
Systematic comparisons through several closely competing polytypes,
in view of their relative stability or details of their band structures, are missing.
($ii$) The issues of band gap are usually treated under the angle of how 
one or another scheme improves its value over the ``conventional'' LDA or GGA
predictions; the modifications of the band structure as a whole and,
in particular, a meaningful analysis of the relative performance of 
meta-GGA versus hybrid functionals in this sense, are not known to us.
($iii$) The assessment of DI for the treatment of these layered systems
is still rare and, whenever done (e.g., Ref.~\onlinecite{PRB90-235302,JPhysChemLett6-3098})
is \emph{ad hoc} and not systematic.

\section{Calculation methods, parameters controlling the accuracy, and XC flavors}
\label{sec:method}
\subsection{WIEN2k and VASP; general setup}
An important objective of our study was to access, in a critical discussion,
different levels of ``sophistication'' and accuracy actually available
for the description of weakly bound layered systems.
Technically, we used two different computer codes, WIEN2k\cite{wien2k} and VASP,\cite{vasp}
with -- in part -- overlapping possibilities, that was of advantage for distinguishing
genuine trends from accidental artifacts of calculation. 
WIEN2k is an all-electron code that employs large basis set of plane waves augmented
to numerical functions within atomic spheres; its accuracy is controlled by 
cutoff parameters for basis function and charge density expansions, for which we used
the values {\tt RKMAX} = 9.0 and {\tt GMAX} = 14.0, correspondingly. As the convergence
of results against enhancing these cutoffs is easy to test, the WIEN2k is able to yield,
in technical sense, the ``DFT truth'' (within the restrictions imposed by the particular
choice of the XC potential flavor).
The VASP code,\cite{PRB47-558,PRB54-11169,vasp} while not being an all-electron
one but using the projected augmented-wave (PAW) scheme \cite{PRB50-17953,PRB59-1758}
for treating the core states, has proven its high accuracy and convenience of use
in a huge number of recent applications to quite different systems.\cite{JCompChem29-2044}
The use of VASP for reliable probing of total-energy preferences between
polytypes imposes setting of some calculation parameters to values different from 
the standard (default) ones. The {\tt PREC} tag, responsible for certain cutoffs,
has to be set to the ``Accurate'' level; {\tt ENCUT} (the planewave cutoff for
the basis functions) was set to 500 eV; {\tt EDIFF} (the convergence criterion
for stopping the electronic relaxation) has to be reduced to {\tt 1E-8}
(from the default value of {\tt 1E-4}), and the criterion of the smallness of forces
on atomic relaxation has to be set not higher than {\tt EDIFFG = -0.01} (in eV/{\AA}),
otherwise the structure relaxation results are too unstable.

A calculation parameter important for both WIEN2k and VASP,
that affects the stability (numerical noise) of total energy results,
is the density of the $\mathbf{k}$ mesh used for the Brillouin zone integration. 
The integration as such was performed by the tetrahedron 
method\cite{SSC9-1763,PRB49-16223} in WIEN2k
and by Monkhorst-Pack sampling\cite{PRB13-5188} in VASP.
Either way, for reliable discrimination of polytypes (by their total energies),
it is essential to enhance the $\mathbf{k}$-mesh density until the total energy
\emph{differences} (not the absolute values) get stabilized 
to the accuracy needed for a meaningful comparison 
of polytypes. For the systems in question this amounts to, as will be seen below, 
to energy differences stabilized within ${\approx}0.2$~meV per formula unit. 
The probing of this criterion and the hence emerging $\mathbf{k}$-mesh densities
(numbers of regular divisions along the reciprocal lattice vectors)
are explained in Fig.~1 of Ref.~\onlinecite{PSSB254-1700120}.
Specifically, one had to go at least up to $\sim$16 divisions along the in-plane (long)
reciprocal lattice vectors of hexagonal lattices, in order to stabilize,
at least qualitatively, the relative placement of energy/volume curves of different
polytypes. In VASP, this is corresponding to at least
a $16{\times}16{\times}4$ $\mathbf{k}$-points grid
(see Subsec.~\ref{subsec:dispersion} for details).

The technical implementation of calculations being thus perfectly controllable,
this is the choice of XC ``flavor'' that accounts for the most remarkable differences.
We proceed at the GGA level, with the parametrization after Perdew--Burke--Ernzerhof 
(PBE),\cite{PRL77-3865,PRL78-1396} arguably one of the most broadly used GGA schemes. 
An important simple modi\-fication of the GGA-PBE parametrization,
that affects the formula for the GGA enhancement factor in view of better reproducing
the properties ``in solids and surfaces'' (rather than of atoms / molecules), 
standardly abbreviated as PBEsol, has been suggested in Ref.~\onlinecite{PRL100-136406}.
We used both PBE and PBEsol, available in both WIEN2k and VASP codes,
as standard ``GGA-only'' schemes in our calculations.

A comparison of band structures calculated with WIEN2k and VASP (for $\beta$-GaSe)
can be found in Fig.4.5 of the Srour's thesis.\cite{Juliana_thesis}
The bands are indistinguishable
for visual assessment even as the corresponding calculations have been done
for not identical XC schemes and, correspondingly, for slightly different
optimized crystal structures.  

A deficiency of our calculation setup is the omission of spin-orbit interaction.
It was included in the pioneering \emph{ab initio} work on InSe
by Gomes da Costa,\cite{PRB48-14135} as well as in 
some recent calculations.\cite{PRB90-235302,JChemPhys141-084701,JPhysChemLett6-3098}
The effect of spin-orbit coupling on the band structure, that can be seen
in Fig.~4 of Debbichi \emph{et al.}\cite{JPhysChemLett6-3098} or in Fig.~7-8
of Ghalouci \emph{et al.},\cite{CompMatSci124-62} is small yet appreciable,
especially in lifting some degeneracies. As these effects will likely pronounce 
in the similar way throughout polytypes, we do not expect the qualitative
trends concerning the latters' relative stability to be affected.
As for the estimations of absolute band gap values, a slight
correction following the inclusion of spin-orbit coupling is quite plausible.

\subsection{Inclusion of dispersion interactions}
\label{subsec:dispersion}
The last decade has witnessed tremendous effort in the development of various
correction methods to account for the DI missing 
in conventional Kohn-Sham DFT calculations. One can single out two types of approaches.
The first one applies specific non-local correlation functional that approximately
accounts for dispersion interactions, in the spirit of that originally developed 
by Dion \emph{et al.}\cite{PRL92-246401} and improved in subsequent 
works.\cite{PRB82-081101,PRB83-195131}
The other group of methods encompasses additive correction schemes, in which
dispersion energy is included on top of ``conventional'' DFT results.
These latter methods (a hierarchy of which is briefly addressed below) typically allow
a relatively easy implementation in the codes, without increasing the calculation
time considerably. A number of such schemes are included in the VASP
package.\cite{JPhysChemA114-11814,JChemTheoComp9-4293,JChemPhys141-034114,
JPCM28-045201,JChemTheoComp12-5920}
We note that there is no implicit electron potential, band structures etc. 
associated to these schemes, but just the total energy (elaborated, in some cases, 
to yield corresponding forces). Consequently, the properties affected concern
just the equilibrium geometry.

In the {DFT+D2} approach of Grimme,\cite{JCompChem27-1787}
the dispersion energy results from summing up the two-atom interactions, which scale 
with interatomic distances $R_{ij}$ as ${\sim}(-R_{ij}^{-6})$ and are moreover enveloped by
a (smeared step-like) \emph{damping function} to prevent spurious overbonding 
at small distances (shorter than about the sum of the van der Waals radii of the atoms concerned). 
The related parametrization is phenomenological and element-related.
The {DFT+D3} approach by Grimme \emph{et al.}\cite{JChemPhys132-154104} adds
an interaction term proportional to ${\sim}R_{ij}^{-8}$ and suggests a different choice
of the damping function than that in Ref.~\onlinecite{JCompChem27-1787}.
The scheme marked as {DFT+D3-BJ} corresponds to a subsequent suggestion
by Grimme \emph{et al.}\cite{JCompChem32-1456}
to modify (yet again) the damping function of DFT+D3, following
the reasoning by Johnson and Becke.\cite{JChemPhys123-024101}
Tkatchenko and Scheffler ({TS}, Ref.~\onlinecite{PRL102-073005}) proposed a way 
to calculate the weighting parameters of ${\sim}R_{ij}^{-6}$ interactions ``on the fly'',
taking into account the modifications of the atoms' static polarizabilities in a given
chemical environment. An ambiguity which may herewith arise in separating the combined 
charge density into atom-related contributions is technically removed
using the Hirshfeld atomic partitioning\cite{TheoChimActa44-129} and notably,
as elaborated by Bu{\v{c}}ko \emph{et al.},\cite{JChemTheoComp9-4293,JChemPhys141-034114}
the ``Iterative Hirshfeld partitioning'', earlier proposed 
by Bultinck \emph{et al.},\cite{JChemPhys126-144111} on top of the TS approach.
The resulting scheme, henceforth referred to as {DFT+TS/HI},
was tested to accurately describe the dispersion interactions in both covalent and
ionic systems.\cite{JChemTheoComp9-4293,JChemPhys141-034114}

A bunch of additive schemes, dubbed {MBD} for ``many-body dispersion'', bypass
the refinement of phenomenological parameters and go directly for the results
expected from the behavior of polarizability functions, making use e.g. 
of the adiabatic-connection fluctuation-dissipation theorem --
see Refs.~\onlinecite{PRL108-236402},~\onlinecite{JChemPhys140-18A508} for details. 
In a nutshell,
the long-range part of the electron correlation energy, missing in ``conventional''
DFT schemes, is recovered via inclusion of (long-range) dipole-dipole interactions
between (short-range-screened) atomic polarizabilities, the latter being represented by
those as for quantum harmonic oscillators. The practical implementation and corresponding
tests (within the VASP code) are described by Bu{\v{c}}ko \emph{et al.}\cite{JPCM28-045201}
An attempt to generalize atomic-related polarizability over the case of variable electron 
number (and hence ionicity), discussed by Gould \emph{et al.},\cite{JChemTheoComp12-5920}
led to a demonstration that the polarizability is piecewise linear in the electron
number, and resulted in corresponding refinement of the MBD scheme. The technical
details related to realization and tests within VASP of this scheme labeled
{MBD/FI} (for Fractional Ions) are given in Ref.~\onlinecite{JChemTheoComp12-5920}.
Note that practical calculations with VASP using the MBD and MBD/FI schemes require,
for maintaining the necessary stability of results, to use much more dense
$\mathbf{k}$-mesh (e.g., 32$\times$32$\times$8 in our case) than 
usual.\cite{JPCM28-045201,JChemTheoComp12-5920}

The WIEN2k code allows the use of D3 corrections 
after Grimme \emph{et al.}\cite{JChemPhys132-154104}
via inclusion of an auxiliary code; otherwise, non-local corrections are implemented
after the scheme by Dion \emph{et al.}\cite{PRL92-246401} and following
the ``efficient implementation'' by Rom{\'a}n-P{\'e}rez and Soler,\cite{PRL103-096102}
the details of which, in what concerns the implementation in WIEN2k and extensive
tests e.g. against VASP, are explained by Tran \emph{et al.}\cite{PRB96-054103}
A number of non-local kernels is provided in WIEN2k.
For practical reasons, we did all the tests concerning the inclusion of vdW
interactions in VASP. 

\subsection{Hybrid functional (HSE06)}
The hybrid XC functionals replace some part of the DFT exchange energy
by the exact exchange from a Hartree-Fock (HF) calculation; this typically
has a favorable effect on the accuracy in prediction of equilibrium geometries; 
moreover the band gap (underestimated in conventional DFT, overestimated in 
HF calculations) becomes closer to reality.
In the present work, we applied the Heyd-Scuseria-Ernzerhof (HSE)
version of a hybrid XC functional,\cite{JChemPhys118-8207,JChemPhys124-219906}
as implemented in the VASP code with modifications of the screening
parameters explained in Ref.~\onlinecite{JChemPhys125-224106}
and casted under the label `HSE06'.
These calculations being relatively time-consuming, we did not
perform full structure relaxation within this scheme, but refer
to HSE band structures, calculated for PBEsol optimized geometry,
for discussion on band gaps and comparison with the mBJ.

\subsection{Modified Becke -- Johnson XC potential}
A technically simple scheme specifically aimed at ``improving'' electron bands
and band gaps via using a particular meta-GGA XC potential have been introduced  
by Tran and Blaha\cite{PRL102-226401} under the name ``modified Becke -- Johnson'' (mBJ),
in the development of the latter authors' idea\cite{JChemPhys124-221101} to explicitly
use the gradient of the kinetic energy density to imitate the characteristic shell
structure of exchange potential in atoms and hence (implicitly) a discontinuity
of the total energy variation with the electron number, a crucial element in
a correct assessment of the band gap.
The implementation in WIEN2k and related extensive tests were described by
Koller \emph{et al.}\cite{PRB83-195134,PRB85-155109}. We note that mBJ is not
a stand-alone total energy functional but just a suggestion for XC potential,
that leads to electron bands but not to total energy / forces.
Corresponding calculations have been performed for
the PBEsol optimized geometry, and comparison done with the HSE band structures.

\begin{figure*}[th!] 
\centerline{\includegraphics[height=0.75\linewidth,angle=-90]{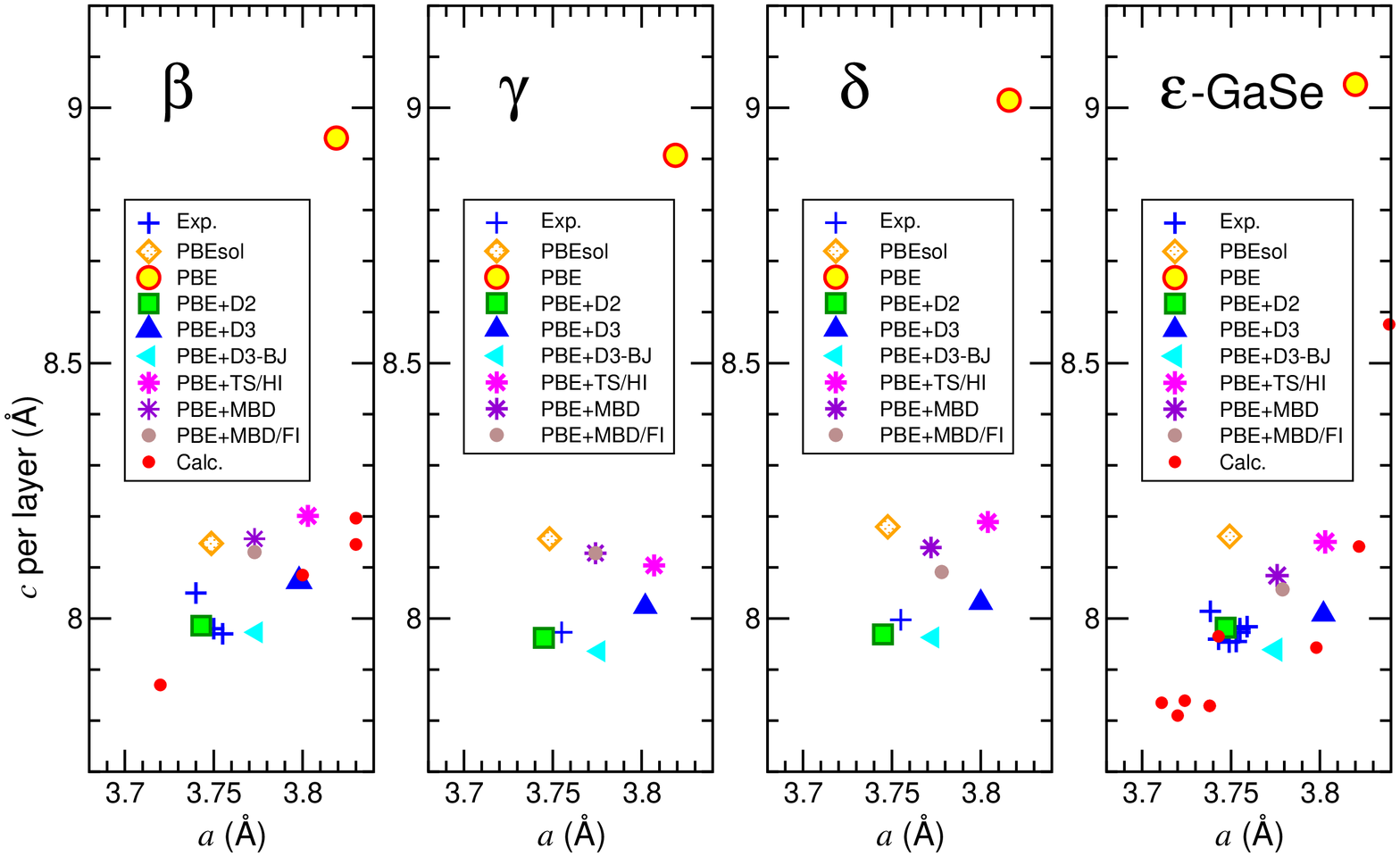}}
\caption{\label{fig:GaSe_a-c}
Lattice parameters in $\beta$, $\gamma$, $\delta$ and $\varepsilon$ phases of GaSe
from the present calculations using the VASP code and from earlier studies. 
The experimental data are indicated by blue crosses, the results of previous calculations -- 
by red dots. Details of Exp. and Calc. values are given in the Supplemental Material.
The schemes of inclusion the dispersion interactions on top of PBE
are explained in Sec.~\ref{subsec:dispersion}. See text for discussion.}
\end{figure*}  

\begin{figure*}[th!] 
\centerline{\includegraphics[height=0.75\linewidth,angle=-90]{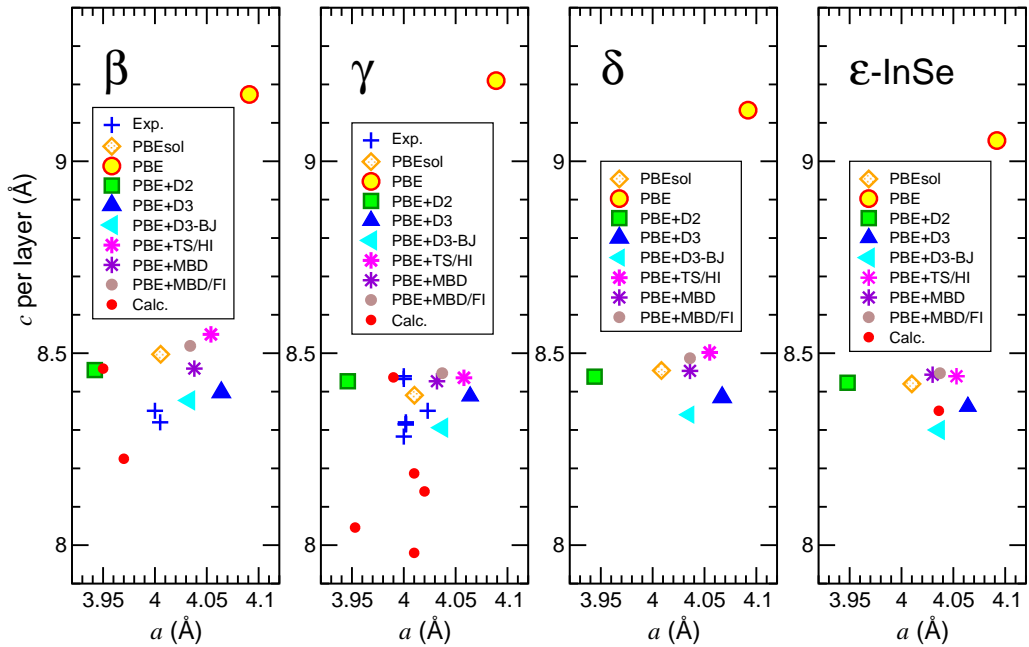}}
\caption{\label{fig:InSe_a-c}
Similar to Fig.~\ref{fig:GaSe_a-c}, for different phases of InSe.}
\end{figure*}   

\section{Optimized crystal structures of different polytypes}
\label{sec:optim}
The lattice parameters as optimized in our calculations, using different XC potentials, 
are indicated in Fig.~\ref{fig:GaSe_a-c} and Fig.~\ref{fig:InSe_a-c}, 
in comparison with earlier calculation results and experiments.
The $c$ parameter, for an easy comparison, is ``normalized'' per one double layer unit
(i.e., divided by two for $\beta$ and $\varepsilon$ phases, by three for $\gamma$,
by four in case of the $\delta$ phase). One can conclude that the standard PBE scheme
systematically overestimates the in-plane $a$ parameter (by $\sim$2.7\%, as compared
with experiment), and especially the $c$ parameter (by $\sim$13\%), that can be primarily
traced to the overestimated interlayer distance. This deficiency is ``pragmatically''
cured by applying (essentially, any) scheme for inclusion of the DI 
in combination with the PBE. Looking more attentively, the performance of PBE+D2
seems to be excellent in GaSe, in comparison with available experimental data;
however, the same scheme yields a too short (by $\sim$2\%) $a$ parameter for InSe.

Similar observations (that PBE gives a wrong $c$ parameter but the equilibrium geometry
can be fixed by the use of PBE+D2) have been done in a recent work
by Rybkovskiy \emph{et al.},\cite{PRB90-235302}
however, without specifying the numerical error nor indicating to which extent
this applies to the three materials studied (GaS, GaSe, InSe).

\begin{table*}[th!] 
\caption{\label{tab:VASPrelax_GaSe}
Optimized crystal structure parameters for different phases of GaSe,
after VASP calculation with PBEsol. $z$-coordinates are formatted such that
they reveal the distances from the double-layer median planes;
see text for details.}
\bigskip
\begin{ruledtabular}
\begin{tabular}{cddcdd}  
\parbox[c]{1.0cm}{Poly-\\*[-2pt]type} & \multicolumn{1}{c}{\hspace*{20pt}$a$~(\AA)}
 & \multicolumn{1}{c}{$c$~(\AA)\rule[-3mm]{0mm}{6mm}} &
\parbox[c]{2.0cm}{Wyckoff\\*[-2pt]position} 
& \multicolumn{1}{c}{\hspace*{18pt}$z_{\rm (Ga)}$} 
& \multicolumn{1}{c}{\hspace*{12pt}$z_{\rm (Se)}$} \\
\hline 
\rule[-4pt]{0mm}{14pt}
$\beta$       & 3.7487 & 2{\times}8.1469 & (f) & \tfrac{1}{4}-(0.14925)/2 & 
\tfrac{3}{4}-(0.29336)/2 \\
\hline 
\rule[0pt]{0mm}{10pt}
$\gamma$ & 3.7482 & 3{\times}8.1563 & (a$^+$) & (0.14905)/3 & \tfrac{2}{3}+(0.29285)/3 \\*[2pt]
\multicolumn{3}{c}{\rule[-4pt]{0mm}{10pt}} & (a$^-$) & -(0.14901)/3 & \tfrac{2}{3}-(0.29290)/3 \\
\hline 
\rule[0pt]{0mm}{10pt}
$\delta$ & 3.7477 & 4{\times}8.1795 & (a$^+$) & (0.14781)/4 & \tfrac{1}{4}+(0.29298)/4 \\*[1pt]
\multicolumn{3}{c}{} & (a$^-$) &               -(0.14914)/4 & \tfrac{1}{4}-(0.29162)/4 \\*[1pt]
\multicolumn{3}{c}{} & (b$^+$) &   \tfrac{1}{4}+(0.14928)/4 & \tfrac{1}{2}+(0.29134)/4 \\*[1pt]
\multicolumn{3}{c}{\rule[-4pt]{0mm}{10pt}} & (b$^-$) &  
\tfrac{1}{4}-(0.14882)/4 & \tfrac{1}{2}-(0.29265)/4 \\
\hline 
\rule[0pt]{0mm}{10pt}
$\varepsilon$ & 3.7492 & 2{\times}8.1610 & 
                       (g) & (0.14897)/2 & \tfrac{1}{2}+(0.29278)/2 \\*[1pt]
\multicolumn{3}{c}{} & (h) & \tfrac{1}{2}+(0.14898)/2 & \multicolumn{1}{c}{} \\*[1pt]
\multicolumn{3}{c}{\rule[-3pt]{0mm}{9pt}} & (i) & \multicolumn{1}{c}{} & (0.29280)/2 \\
\end{tabular}
\end{ruledtabular}
\end{table*} 

\begin{table*} 
\caption{\label{tab:VASPrelax_InSe}
Similar to Table~\ref{tab:VASPrelax_GaSe}, for InSe.}
\bigskip
\begin{ruledtabular}
\begin{tabular}{cddcdd}  
\parbox[c]{1.0cm}{Poly-\\*[-2pt]type} & \multicolumn{1}{c}{\hspace*{20pt}$a$~(\AA)} & 
\multicolumn{1}{c}{$c$~(\AA)\rule[-3mm]{0mm}{6mm}} &
\parbox[c]{2.0cm}{Wyckoff\\*[-2pt]position} 
& \multicolumn{1}{c}{\hspace*{18pt}$z_{\rm (In)}$} 
& \multicolumn{1}{c}{\hspace*{12pt}$z_{\rm (Se)}$} \\
\hline 
\rule[-4pt]{0mm}{14pt}
$\beta$       & 4.0055 & 2{\times}8.4972 & (f) & \tfrac{1}{4}-(0.16394)/2 & 
\tfrac{3}{4}-(0.31429)/2 \\
\hline 
\rule[0pt]{0mm}{10pt}
$\gamma$ & 4.0102 & 3{\times}8.3906 & (a$^+$) & (0.16553)/3 & \tfrac{2}{3}+(0.31796)/3 \\*[1pt]
\multicolumn{3}{c}{} & (a$^-$) &               -(0.16552)/3 & \tfrac{2}{3}-(0.31798)/3 \\
\hline 
\rule[0pt]{0mm}{10pt}
$\delta$ & 4.0085 & 4{\times}8.4550 & (a$^+$) & (0.16549)/4 & \tfrac{1}{4}+(0.31437)/4 \\*[1pt]
\multicolumn{3}{c}{} & (a$^-$) &               -(0.16286)/4 & \tfrac{1}{4}-(0.31696)/4 \\*[1pt]
\multicolumn{3}{c}{} & (b$^+$) &   \tfrac{1}{4}+(0.16290)/4 & \tfrac{1}{2}+(0.31691)/4 \\*[1pt]
\multicolumn{3}{c}{\rule[-4pt]{0mm}{10pt}} & (b$^-$) &   \tfrac{1}{4}-(0.16552)/4 & \tfrac{1}{2}-(0.31433)/4 \\
\hline 
\rule[0pt]{0mm}{10pt}
$\varepsilon$ & 4.0100 & 2{\times}8.4201 & 
                       (g) & (0.16492)/2 & \tfrac{1}{2}+(0.31680)/2 \\*[1pt]
\multicolumn{3}{c}{} & (h) & \tfrac{1}{2}+(0.16488)/2 & \multicolumn{1}{c}{} \\*[1pt]
\multicolumn{3}{c}{\rule[-3pt]{0mm}{9pt}} & (i) & \multicolumn{1}{c}{} & (0.31686)/2 \\
\end{tabular}
\end{ruledtabular}
\end{table*} 

Alternatively and pragmatically, the PBEsol XC scheme seems to
perform quite well without any additional inclusion of the DI. On the contrary,
the combination of PBEsol with the D3 and moreover the BJ schemes tends to overbind
too much, resulting in underestimation of both $a$ and $c$ (by $\sim$1$-$2\%;
the corresponding data are not included in Fig.~\ref{fig:GaSe_a-c},\ref{fig:InSe_a-c},
but can be found in Fig.~4.1 of the Srour's thesis.\cite{Juliana_thesis} This observation
holds for both GaSe and InSe systems and, within some data scattering, throughout
all phases for which the experimental parameters are available. 
Rybkovskiy \emph{et al.}\cite{PRB90-235302} report that PBEsol improves the 
in-plane distances, in comparison to PBE, but still overestimates the interlayer
separation (without further elaborating).

Discarding PBE+D2 as not sufficiently reliable (at least for InSe), we can mark a fair 
agreement within the other (``better'') schemes of including the DI 
on top of PBE. Somehow comfortingly, the schemes
which are a priori expected to be more accurate and flexible do indeed yield 
more accurate prediction of the lattice
parameters. Even as experimental lattice parameters are available for some phases only
($\beta$, $\varepsilon$-GaSe and $\beta$, $\gamma$-InSe), the stability
of $a$ and ``reduced'' $c$ throughout polytypes seems plausible. In this perspective,
an accurate (and consistent) performance of PBE+D3-BJ, arguably the best among
``phenomenological'' schemes, and more sophisticated TS/HI and MBD approaches
seem reassuring. One can note not much difference in the MBD results with and
without ``fractional ions'' modification, for an apparent reason that our materials
do not possess a strongly ionic character. We'll come to the differences
in the energy / volume curves yielded by different dispersion schemes in the next section.

The details of the crystal structure, optimized throughout phases with ultimate
$\mathbf{k}$-mesh of (24$\times$24$\times$6), are given in Tab.~\ref{tab:VASPrelax_GaSe}
and \ref{tab:VASPrelax_InSe} (from PBEsol calculations only).
The internal coordinates are expressed in such way as to facilitate their comparison
throughout polytypes; namely, the values in the numerator (e.g.,
$\simeq\,$0.15 for Ga, $\simeq\,$0.29280 for Se) everywhere play a role of
deviation from the median plane of a double layer, in units of ``reduced''
$c$ parameter (e.g., $\simeq\,$8.2~{\AA} for GaSe). 
Note that in $\gamma$ and $\delta$ phases, the median planes of double layers
are not fixed by symmetry, and an arbitrary rigid shift of all the $z$ coordinates can be applied.
Our $z$ scale were in these cases gauged so as to ``equilibrate'' 
positive and negative deviations for all the ions.
With this, the $\gamma$ phase (of both GaSe and InSe) maintains its
double layers practically mirror symmetric with respect to the median plane.
For the $\delta$ phase, on the contrary, the $(+)$ and $(-)$ coordinates within either (a) or (b)
positions are \emph{not} symmetric, meaning that each double layer is polarized
up or down; however, there is an approximate criss-cross symmetry between (a) and (b)
positions, so that the (a$^{+}$) distance from the median plane nearly equals that for (b$^{-}$),
and vice versa, for a given atom species. This means that the up / down polarizations
of consecutive double layers are alternating throughout the stacking. This disparity
of symmetry-breaking alternating displacements (of the order of 1\%), more pronounced for InSe
than in GaSe, is schematically shown in Fig.~2.5 of Ref.~\onlinecite{Juliana_thesis}.
Apart from this systematic ``flaw'', the coordinates remain remarkably stable over
the polytypes; the variations throughout the phases of InSe
($\lesssim\,$1.2\% over $c$ values, $\lesssim\,$0.7\% over cation-cation distances)
are just minutely more pronounced than in the case of GaSe 
($\lesssim\,$0.4\% and $\lesssim\,$0.1\%, correspondingly).

Obviously, the accuracy in absolute values suggested by Tables~\ref{tab:VASPrelax_GaSe}, 
\ref{tab:VASPrelax_InSe}
exceeds by far the credibility of contemporary first-principles schemes; 
nevertheless, the systematic errors are likely to be common for different polytypes,
so that the qualitative trends should presumably hold. 
More instructive than
just the equilibrium geometries are the energy profiles around the corresponding minima,
discussed in the following section.

\begin{figure}[h!] 
\includegraphics[width=1.0\linewidth]{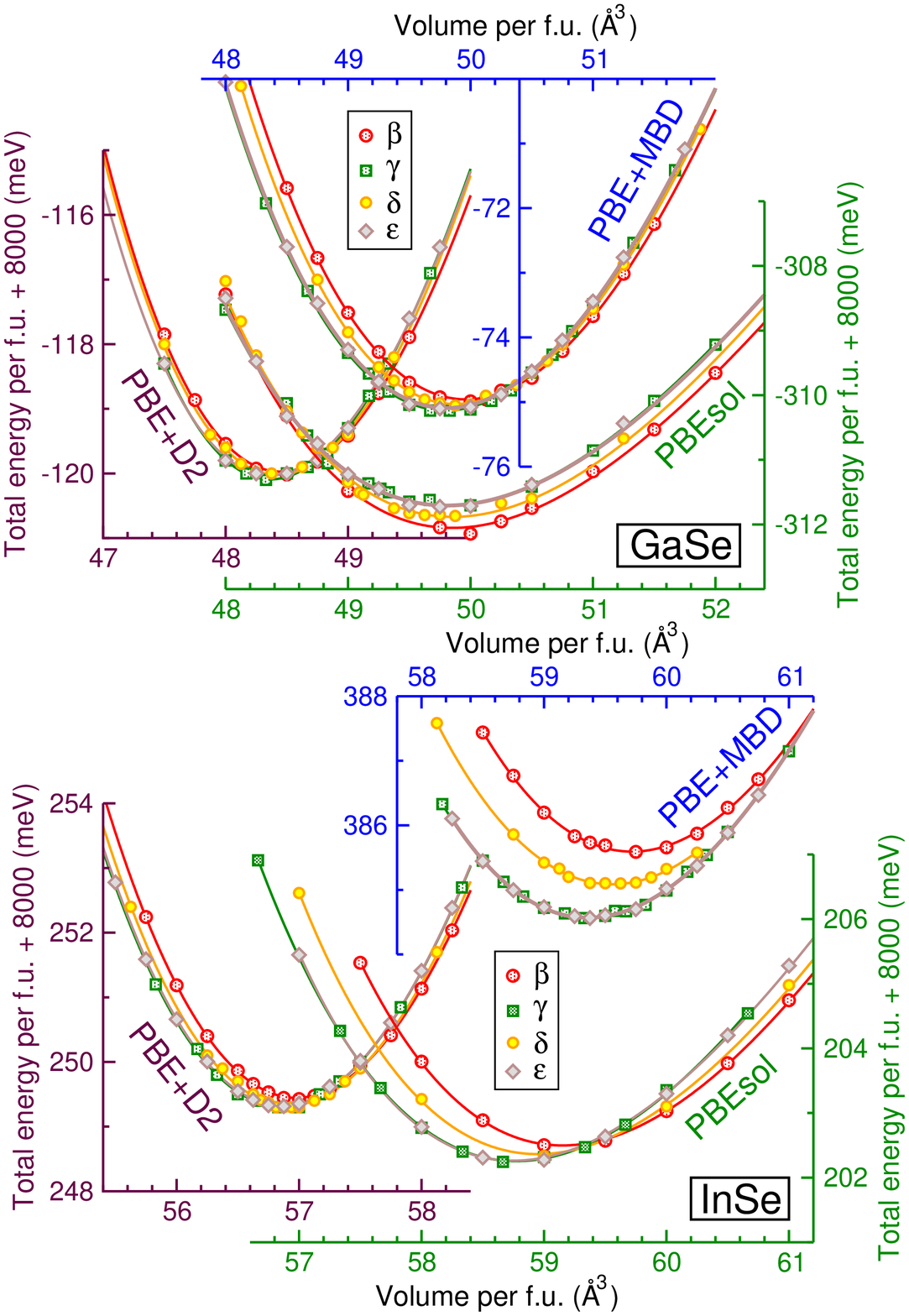}
\caption{\label{fig:EneVol}
Energy / volume curves for four polytypes of GaSe (upper panel) and InSe (lower panel),
as calculated with PBEsol (a bunch of curves at bottom-right of each panel),
with dispersion interaction included according to the D2 Grimme scheme
on top of PBE (a bunch of curves at the left of each panel)
and with many-body dispersion interaction included on top of PBE
(a bunch of curves at the top of each panel).
Note that the volume axis and the relative energy scale are common throughout each panel,
whereas absolute energy values are unrelated. 
Symbols indicate total energies after full structure relaxation for a given volume;
the lines are the Murnaghan fit through these data. See text for discussion.}
\end{figure}   

\section{Equations of state and relative stability of polytypes}
\label{sec:eq_state}
The energy / volume $E(V)$ curves for different polytypes have been earlier studied  
in Ref.~\onlinecite{PSSB254-1700120}, using the PBEsol XC potential and the
WIEN2k vs VASP calculation methods in comparison. 
In addition to a nominal result of which phase has lower energy at equilibrium,
the curve as a whole indicates how the relative stability of phases
would be shifted by (positive or negative) pressure.
Since the $E(V)$ curves for different polytypes are (near the respective minima)
close within 1~meV per double unit (i.e., 4 atoms -- see Fig.~2 and 3
of Ref.~\onlinecite{PSSB254-1700120}),
a reliable (noise-free) resolving them required an utmost
care in controlling the numerical accuracy (in terms of $\mathbf{k}$-mesh, see Fig.~1
of Ref.~\onlinecite{PSSB254-1700120}, and planewave cutoffs). 
In the present study, we focus at the effects of including the DI,
in the form of a simple Grimme D2 scheme and with more sophisticated MBD,
both on top of the PBE XC potential; the results are depicted in Fig.~\ref{fig:EneVol}.
Every point in the curves corresponds to a full relaxation of $a$, $c$ and internal
coordinates for a given trial volume. Fitting to the Murnaghan equation of state yields
the bulk moduli, shown in Table~\ref{tab:bulk_mod} and discussed further on.

We note that the volume axes in Fig.~\ref{fig:EneVol} are consistent throughout
the three panels shown for each compound, whereas the total energy values 
from different methods are obviously unrelated, and superposed arbitrarily.
The absolute energy values are indicated just for reference.

Two observations can be done concerning the general ``impression'' of
Fig.~\ref{fig:EneVol}:
$(i)$ For both compounds and all the calculation schemes, the $E(V)$ curves
for $\gamma$ and $\varepsilon$ polytypes stay practically degenerate,
within the meaningful accuracy, in spite of their technically
not identical treatment (cell size, exact $\mathbf{k}$-mesh);
$(ii)$ this merged ($\gamma$, $\varepsilon$) curve is in all cases the most 
distinct from that for the $\beta$ polytype, the $\delta$ curve taking its position
cleanly in between.
This can be understood from the differences in the double-layer packing, or, specifically,
how the next layer is placed on top of the previous one. The $\beta$ phase is characterized
by a ``double lock'' whereby the next-layer anion sits on top of the current-layer cation
and vice versa (in the eclipsed configuration),
this schema going on in both senses (the [Bc~Cb] packing, see Fig.~\ref{fig:InSe_struc}).
In $\gamma$ and $\varepsilon$, the next-layer cation is placed on top of the current-layer anion;
however, the next double layer is pivoted and the reverse cation-anion ``lock'' across
the interlayer gap is missing. The $\gamma$ [Ca~Ab~Bc] and the $\varepsilon$ [Ca~Ab] phases
differ only in what concerns the packing beyond the nearest-neighboring double layer. 
Under this angle, the $\delta$ phase is indeed intermediate:
out of its four double layers, two and the next two are pairwisely in ``double lock'', 
with pivoted ``loose locks'' in between. (This also explains the above discussed
asymmetry / alternating polarization of double layers in the $\delta$ phase).
A conclusion from this analysis is that, since all other imaginable polytypes
cannot be but various combinations of ``double locked'' and ``loosely locked''
double layers, their relative $E(V)$ curves are very likely to fall
between the limits drawn by $\beta$ and ($\gamma$ or $\varepsilon$) ones.

Addressing the issue of energy preference of different polytypes,
one can note that the situation (for both compounds and all calculation methods)
is not such that one of the above ``limiting'' curves would fully encompass
the other and definitely ``win''.\footnote{%
This is at variance with the results by Ghalouci \emph{et al.}\cite{CompMatSci124-62}
for InSe, obtained with the GGA, according to which the $\beta$-InSe is lower in energy
than $\varepsilon$-InSe by about 100~meV per unit cell.}
Rather, the curves are crossing not far from
their respective minima, so that the equilibrium in favor of one or the other phase
is likely to be shifted
under a moderate effect of pressure. In all cases, the $\beta$ phase would eventually win
at large enough volume (hence negative pressure), and $\gamma$ / $\varepsilon$ --
under positive pressure, from small enough volume downwards. As it turns out from the results of
PBEsol calculations, GaSe definitely prefers the $\beta$ phase at the ambient conditions,
and needs the pressure of $\simeq\,$0.6~GPa (judging by the common tangent, to be drawn on the left,
where the curves cross) to be pushed into the $\gamma$ or $\varepsilon$ phase.
InSe, on the contrary, tends for $\gamma$ / $\varepsilon$ phase
at zero pressure, but a small expansion (negative pressure of $-$0.1~GPa)
would make the $\beta$ phase competitive. We note in this relation that rhombohedral ($\gamma$)
was, indeed, early enough identified as \emph{the} structure of single-crystal 
InSe,\cite{ComRenAcadSciC279-33} whereby even earlier reports (by Semiletov, in 1958)
of detecting a hexagonal two-layer phase were attributed in Ref.~\onlinecite{ComRenAcadSciC279-33}
to ``very peculiar conditions'' (by evaporation in vacuum) of preparing
the thin-film samples in question.

The $E(V)$ profiles calculated in PBE+D2 are markedly contracted
(see the increased values of the bulk moduli in Table~\ref{tab:bulk_mod}),
to the point that
different polytypes become almost indistinguishable in the scale of Fig.~\ref{fig:EneVol}.
This holds for both GaSe and InSe.
An explanation could be that the D2 scheme, only sensitive to interatomic distances
but not to details of the short-range arrangement of atoms, is too crude 
to make distinction between the polytypes.
The shift of the curve to smaller volumes with simultaneous increase of its stiffness
due to inclusion of the D2 correction is generally known (see, e.g., Fig.~1 of
Ref.~\onlinecite{JPhysChemA114-11814}).

The PBE+MBD calculation, sensitive to the charge density distribution,
recovers the discrimination between polytypes, 
to the effect that is differently pronounced in GaSe and InSe. In InSe,
the sequence of phases is roughly the same as after the 
PBEsol calculation (the $\gamma$ or $\varepsilon$ phases are dominating at ambient
conditions and up to appreciable negative pressure); the stiffness is slightly
larger than that estimated by PBEsol. In GaSe, the ground-state phase
at ambient pressure, by very small margin,
according to PBE-MBD calculation would be $\gamma$ or $\varepsilon$;
the preference of the $\beta$ phase can be restored by negative pressure of $-$0.2~GPa.

In total, the hierarchy of phases in InSe seems relatively robust as different
calculation schemes are applied; the whole set of curves just gets uniformly
compressed and slightly shifted. This could be related to relatively higher
covalence of InSe, whereby the interlayer interactions are to some extent
already grasped within the conventional DFT, different polytypes are reliably
discriminated on the basis of the (small as it is) covalent part in their interlayer 
coupling, and ``perturbations'' due to different ways of including dispersion interactions 
do not change the qualitative trend. In GaSe, the ``conventional'' chemical bonding
is to larger extent confined within the double layer, so that the role of 
dispersion interactions (and, consequently, of the diversity in their practical inclusion)
comes out more pronounced.

\begin{table*} 
\caption{\label{tab:bulk_mod}
Calculated values of the bulk modulus $B_0$ and its pressure derivative $B'$
for GaSe and InSe in comparison with available experimental and calculation results. 
The span of values for $B_0$ covers different polytypes. $B_{\rm R}$, $B_{\rm V}$ 
stand for the Reuss average (lower bound) and Voigt average (upper bound) 
of the bulk modulus estimated from the elastic constants -- see text for detail.}
\bigskip
\begin{ruledtabular}
\begin{tabular}{lcccc}  
 & \multicolumn{2}{c}{GaSe} & \multicolumn{2}{c}{InSe} \\
  \cline{2-3}  \cline{4-5}
  \rule[-2pt]{0mm}{12pt}
Method & $B_0$ (GPa) & $B'$ & $B_0$ (GPa) & $B'$ \\
\hline 
PBEsol  & {$11.1 - 12.1$} & {$21 - 31$} & {$15.0 - 15.6$} & {$23 - 25$}     \\
PBE+D2  & {$32.7 - 33.7$} & {$10 - 14$} & {$31.0 - 31.5$} & {$5 - 10$} \\
PBE+MBD & {$20.4 - 21.3$} & {$13 - 16$} & {$20.9 - 21.6$} & {$12 - 16$} \\
\hline 
expt. volume (pressure) fit  \rule[-2pt]{0mm}{11pt} & 34(2)$^{a}$ & 6.4(5)$^{a}$ 
\\
\parbox[l]{5.0cm}{$B_{\rm R}\,\cdots\,B_{\rm V}${\hfill}~ \\*[-1pt] 
(from expt. elastic constants){\hfill}~} &
\parbox[c]{4.0cm}{$\left\{\begin{array}{c}
27.7\,\cdots\,38.5^{b} \\ 28.6\,\cdots\,38.8^{c} \end{array}\right.$} & & $34.6\,\cdots\,39.6^{b}$
\\
\parbox[c]{5.0cm}{calc. energy (volume) fit {\hfill}~\\*[-1pt] 
(two choices of $B'$){\hfill}~} & & & 
\parbox[c]{1.5cm}{$\left\{\begin{array}{c} 29^{d} \\ 34^{d} \end{array}\right.$} &
\parbox[c]{1.5cm}{$\begin{array}{c} 6.2^{d} \\ 5^{d}\; \end{array}$} 
\\
\parbox[l]{5.0cm}{$B_{\rm R}\,\cdots\,B_{\rm V}${\hfill}~ \\*[-1pt] 
(from calc. elastic constants){\hfill}~} &
\parbox[c]{4.0cm}{$\left\{\begin{array}{c}
28.4\,\cdots\,38.4^{e} \\ 28.3\,\cdots\,38.2^{f} \end{array}\right.$}
\\
\end{tabular}
\end{ruledtabular}
\smallskip
$^{a}$Ref.~\onlinecite{PSSB244-244}; $^{b}$Ref.~\onlinecite{PSSB119-327};
$^{c}$Ref.~\onlinecite{JPCM11-6661}; $^{d}$Ref.~\onlinecite{PhysSolState46-179};
$^{e}$Ref.~\onlinecite{PRB57-3726};  $^{f}$Ref.~\onlinecite{ChinPhysLett23-1876}.
\end{table*} 

The calculated values of the bulk moduli $B_0$ in Table~\ref{tab:bulk_mod}
offer another interesting benchmark  
concerning the performance of different calculation schemes.
We see that PBEsol, PBE+D2 and PBE+MBD yield three groups of $B_0$ values
($\simeq$11$-$15~GPa, $\simeq$32~GPa and $\simeq$21~GPa, respectively),  
whereby the differences between polytypes within each group are comparable  
with technical errors of fitting (depending on the range chosen, etc.).
According to PBEsol calculation, GaSe comes out noticeably softer than InSe;
however, each of two other calculation schemes yields very close (within $\simeq$6\%)
values of $B_0$ for GaSe and InSe. The experimental estimations of
the bulk moduli of the two crystals, in view of the scattering of the data reported,
look indeed quite identical.
In addition to (rarely) reported face values of bulk modulus $B_0$ along with its
pressure derivative $B'$, some previous works listed the elastic constants,
from which the Reuss average and the Voigt average\footnote{%
$B_{\mbox{\tiny [Reuss]}}=
[(C_{11}\!+\!C_{12})C_{33}-2C_{13}^2]/[C_{11}\!+\!C_{12}\!+\!2C_{33}-4C_{13}]$;
$B_{\mbox{\tiny [Voigt]}}=[2(C_{11}\!+\!C_{12})\!+\!4C_{13}\!+\!C_{33}]/9$.}
can be extracted, known to be correspondingly the lower and the upper bound
for $B_0$.
One notes that the hardening of the $E(V)$ profile with the use of PBE+D2 scheme
almost ideally reproduces the experimental values; the PBEsol results
without including the DI are markedly ``too soft'', whereas the PBE+MBD scheme,
presumably the most accurate one (among those tested) in the prediction of lattice
parameters, apparently gives a fair yet systematically slightly underestimated
$B_0$ values.

Interestingly, the earlier \emph{ab initio} estimations of 
bulk moduli\cite{PRB57-3726,ChinPhysLett23-1876,PhysSolState46-179}
shown in Table~\ref{tab:bulk_mod} are quite close to experiment. These calculations
have been done with the LDA and hence result in
slight overbinding (that somehow compensates for the missing DI) and in a corresponding
hardening of the $E(V)$ profile to almost exemplary values (albeit for a wrong reason).
It would have been very instructive to probe within the PBE+MBD scheme
the elastic constants separately, in order to find out where the presumed
deficiency of the resulting bulk modulus comes from.

\section{\lowercase{m}BJ- and HSE-corrected band structures and band gaps}
\label{sec:band_gaps}
\subsection{Band foldings in different polytypes}
Energy bands (which are the origin of the total energies and the hence derived
differences between polytypes) are basically formed by interactions within the double layer,
then get replicated and distorted according to how the number of units
varies throughout the polytypes. Fig.~2 of Rybkovskiy \emph{et al.}\cite{PRB84-085314} 
is an instructive example of realistic DFT band structure calculated for an isolated
double layer of GaSe, with its seven valence bands (counting upwards from Ga$4s^2$,
at about 7~eV below the valence band top, followed by Ga$4p$ and Se$4p^4$).
The closest approximation to it in our case is the band structure of $\gamma$-GaSe,
with one double-layer unit per rhombohedral primitive cell.\footnote{An extension
of Ref.~\onlinecite{PRB84-085314} onto 2, 3 and 4 GaSe double layers\cite{JNanoOpto7-65}
provides a didactically nice example of bands' multiplication in a two-dimensional
band structure.} 
In the following figures, the $\mathbf{k}$-path is uniformly chosen in the 
hexagonal setting throughout all the polytypes; the $\gamma$-GaSe would therefore exhibit
three times more bands in the hexagonal setting than in the rhombohedral one
(see the discussion below).
We skip discussion of ``conventional'' GGA bands as not particularly relevant
and go directly for the systematic analysis of ``corrected'' band structures
(within mBJ and HSE) which would also enable us to discuss the band gaps
in comparison with experimental data.

\begin{figure}[h!] 
\includegraphics[width=1.0\linewidth]{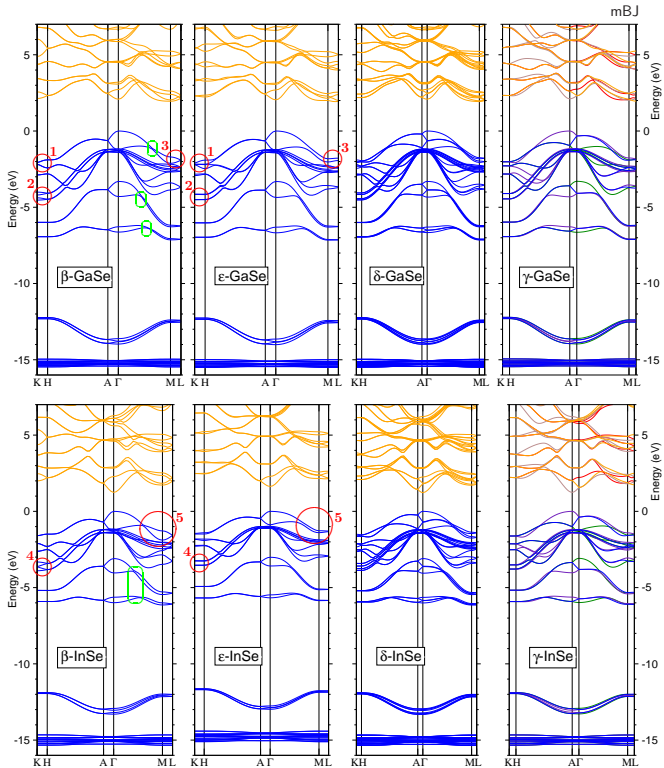}
\caption{\label{fig:mBJ_bands}
Band structures of GaSe (upper row) and InSe (lower row) polytypes
calculated by WIEN2k with mBJ. Occupied bands are drawn in blue, vacant bands -- in yellow; 
for the $\gamma$ polytypes, an additional color coding is used -- see text for discussion.
Zero energy is set at the conduction band top.
Red numbered circles and green ovals indicate the elements addressed in the text.
$A$, $H$ and $L$ points are on top of respectively $\Gamma$, $K$ and $M$.}
\end{figure}  

\subsection{mBJ band structures; differences between polytypes}
The band structures calculated with mBJ are shown in Fig.~\ref{fig:mBJ_bands}.
Some fragments which merit attention and to which the reference is made in the text
are marked by numbered red circles.

The comparison of $\beta$ and $\varepsilon$ phases, which have 
the same number of bands and basically similar dispersions, reveals 
differences in band splittings and degeneracies in some symmetry points,
or along some symmetry lines. Since the structural difference between $\beta$ and 
$\varepsilon$ is in the stacking of otherwise identical layers, the differences
in the band dispersion come about at the BZ boundary, along $K-H$ and $M-L$,
and affect the bands of predominantly Se$4p$ character, the most ``sensitive''
to the mutual orientation of adjacent double layers. The two upper occupied bands 
in the $\beta$ phase proceed as doubly degenerate ones along $K{\rightarrow}H$ 
(and further on towards $A$),
whereas the degeneracy is lifted in their counterparts of the $\varepsilon$ phase
(label 1 in Fig.~\ref{fig:mBJ_bands}). In the ``adjacent part'' of the BZ boundary,
along $M{\rightarrow}L$, the said two upper bands converge towards degeneracy,
whereas in the $\varepsilon$ phase they proceed almost parallel (label 3;
also label 5 in case of InSe).
An opposite pattern of splitting comes about for a pair of lower placed bands
(in the range $-4$ / $-5$ eV in GaSe and $-3$ / $-4$ eV in InSe) which proceed as double degenerate ones
along $H-K$ in $\varepsilon$ but markedly split from $H$ towards $K$ in the $\beta$ phase
(label 2 for GaSe and label 4 for InSe in Fig.~\ref{fig:mBJ_bands}).

Generally, many bands remain doubly degenerate on the upper (flat) BZ boundary,
e.g., along $H$ -- $A$, but get split on going inside ($A$ -- $\Gamma$),
and the degeneracy is lifted in the basal ($\Gamma$ -- $M$) plane. It is noteworthy
how some of these split bands go side by side
in $\varepsilon$ phase but undergo a crossing in the $\beta$ phase. Such crossings
are marked in Fig.~\ref{fig:mBJ_bands} by green ovals. 

Large band dispersions along $\Gamma$ -- $A$ reveal the interaction between the
double layers. On passing from two to four double layers, the BZ is halved and the 
$A$ -- $H$ path is backfolded onto $\Gamma$ -- $K$ etc., doubling the number of 
bands. This becomes obvious from comparing the band structures of the 
(four double layers) $\delta$ phase with those of (two-layers) $\beta$ or $\varepsilon$. 
In case of the $\gamma$ polytype, the situation is more delicate. The primitive cell
is rhombohedral; plotting the band structure in the hexagonal setting amounts
to superposing three band structures calculated along three $\mathbf{k}$ paths,
the original one and the two displaced by ${\pm}1/3$ of the BZ height. 
Such band structures are marked in the
right-hand side panels of Fig.~\ref{fig:mBJ_bands} by different colors,
separately for occupied and unoccupied bands.\footnote{The explicit superposition
of three ``partial'' band structures, calculated with PBEsol for $\gamma$-GaSe,
can be found in Fig.~4.4 of Ref.~\onlinecite{Juliana_thesis}.}
The most spectacular consequence from the fact that the $\gamma$ phase possesses
an odd number of double layers is that, due to a forth-back-forth folding of bands,
the valence-band top and the conduction-band bottom
occur not in $\Gamma$ but in $A$.

As is well seen from the GaSe band structures in Fig.~\ref{fig:mBJ_bands}, the local minimum
of the valence band in $M$ competes with that in $\Gamma$ for being the global one.
Should this happen, the band gap would become indirect, since the
valence band top remains always in $\Gamma$. The band gap values for direct and indirect
gaps are given in Table~\ref{tab:gaps}, to be discussed below. For InSe, the band gap
is direct for all polytypes.

All these observations are not necessarily specific to mBJ, but we discuss them
since we consider the mBJ band structure reasonably accurate in the absence of 
superior quality (say, GW) calculations. Moreover we'd like to emphasize that
the preferences in total energies between polytypes are not accidental but may,
in principle, be traced down to particular features in the band dispersions.

\begin{figure}[t!] 
\includegraphics[width=1.0\linewidth]{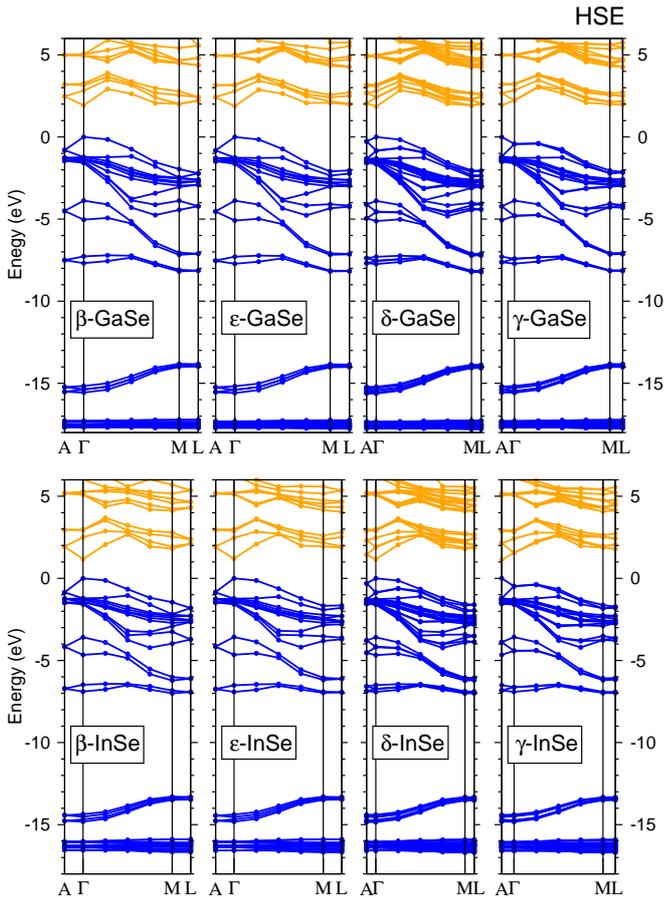}
\caption{\label{fig:HSE_bands}
Band structures of GaSe (upper row) and InSe (lower row) polytypes,
calculated by VASP with HSE06 (for few $\mathbf{k}$-points only along the path).
The color coding and setting the energy zero are as in Fig.~\ref{fig:mBJ_bands}.
}
\end{figure} 

\subsection{HSE band structures}
The hybrid-functional calculations has become another broadly accepted way 
to ``improve'' the underestimated band gaps, along with the band structure as a whole,
as compared to ``conventional'' DFT calculation. Band structures calculated with HSE06
along (in part) the same $\mathbf{k}$ path as that is Fig.~\ref{fig:mBJ_bands} 
are shown in Fig.~\ref{fig:HSE_bands}.
The distribution of $\mathbf{k}$ points along the path is relatively sparse,
as compared to quasi continuous one in Fig.~\ref{fig:mBJ_bands}. In fact, as hybrid calculations
are relatively costly, the $\mathbf{k}$ points from 
the regular grid used for the BZ integration have been selected. As we do not discuss
the total energies extracted from hybrid-functional calculations, 
for purely illustrative purposes such sparse grid seems to be acceptable. We can confirm
that all the above observations concerning the band splittings in $\beta$ vs $\varepsilon$ phases
(red circles in Fig.~\ref{fig:mBJ_bands}) remain valid for HSE calculations.
The remarks concerning the band crossings are not conclusive here, due to a sparseness
of the $\mathbf{k}$ grid. The quantitative differences in the band gap values
are discussed in the following.

\subsection{General observations from comparing the mBJ and HSE band structures}
Whereas the need for reliable band gap predictions remains probably the major
motivation behind using these schemes, one should not overlook that in the process
the whole band structure gets somehow, and differently, modified, as compared
to ``conventional'' DFT predictions. This reveals the fact that differently localized
states are differently affected by the two formalisms. The HSE inherits from the 
Hartree-Fock the tendency to place occupied (e.g., semicore) states too low.
Taking the valence band top for zero energy, we find (in $\beta$-GaSe) 
the bottom of the Se$4p$-related valence band at ${\sim}-6.2$~eV with mBJ
(roughly the same as with PBEsol, see Fig.~4.3 of Ref.~\onlinecite{Juliana_thesis}) 
but at ${\sim}-7$~eV with HSE; 
the Ga$4s$-related flat band that spans ${\sim}-7$ to $-6$~eV
in PBEsol and mBJ shifts by ${\sim}1$~eV downwards with HSE;
the Se$4s$-related bands that span ${\sim}-14$ to $-12.5$~eV 
plunges down by ${\sim}1.5$~eV, and the bunch of flat Ga$3d$ at ${\sim}-15.2$~eV
are found more than 2~eV deeper in a HSE calculation than in mBJ.
Interestingly, in InSe the plunging of these semicore bands is less spectacular;
notably the In$4d$ is deepened (in HSE, as compared to mBJ) by ${\sim}1.5$~eV only,
apparently due to a weaker localization of these states as compared to Ga$3d$.
One can note that the mBJ calculation predicts the ``gap'' between the two lowest
bunches of bands included in the figures, i.e., the Se$4s$ and the cation-$d$,
smaller in GaSe than in InSe, whereas this is the other way around with HSE.
It could be instructive to resort to electron spectroscopy studies, which we failed
to find in earlier publications, for a critical assessment of the semicore bands' placement
according to mBJ and HSE formalisms.

\begin{table*}[th!] 
\caption{\label{tab:gaps}Calculated band gap values (in eV) for different polytypes 
of GaSe and InSe in comparison with experiments and available GW calculations} 
\medskip
\begin{ruledtabular}
\begin{tabular}{lcdddd}  
  & & \multicolumn{4}{c}{polytypes} \\
  \cline{3-6}
  Method & \parbox[b]{2.0cm}{gap\\*[-1pt] nature} 
  & \multicolumn{1}{c}{$\beta (2H)$}  
  & \multicolumn{1}{c}{$\gamma (3R)$} 
  & \multicolumn{1}{c}{$\varepsilon (2H)$} 
  & \multicolumn{1}{c}{$\delta (4H)$} \\
\hline & \rule[0pt]{0mm}{10pt} & \multicolumn{4}{c}{GaSe} \\
PBEsol       & (direct):   & 0.934^{\dag}  & 0.924^{\S} & 0.745^{\dag}  & 0.853^{\dag}  \\
mBJ          & (direct):   & 2.092^{\dag}  & 2.113^{\S} & 1.889^{\dag}  & 2.010^{\dag}  \\*[-1pt]
HSE          & (direct):   & 1.928^{\dag}  & 1.931^{\S} & 1.881^{\dag}  & 1.856^{\dag}  \\
mBJ          & (indirect): & 1.949^{\ddag} & 1.963^{\P} & 1.786^{\ddag} & 1.886^{\ddag} \\*[-1pt]
HSE          & (indirect): & 2.219^{\ddag} & 1.971^{\P} & 1.976^{\ddag} & 2.014^{\ddag} \\
\cline{4-5}
Exp.         & (direct):   & 2.169^a & \multicolumn{2}{c}{2.120$^a$; 2.0196$^b$} \\*[-0pt]
Exp.         & (indirect): & 2.117^a & \multicolumn{2}{c}{2.065$^a$; 2.010$^b$}  \\
Exp.         & (direct)    &         &      & 2.020^c \\*[-1pt]
Exp.         & (indirect)  &         &      & 1.995^c \\ 
Exp.    & (exciton peaks): & 2.050^d &      & 2.004^d & 2.026^d \\  
Calc. GW     & (direct)    &         &      & \multicolumn{1}{c}{2.34$^e$; 1.75$^f$; 2.11$^g$} \\
\hline & \rule[0pt]{0mm}{10pt} & \multicolumn{4}{c}{InSe} \\
PBEsol       & (direct):   & 0.304^{\dag}  & 0.240^{\S} & 0.731^{\dag}  & 0.607^{\dag}  \\
mBJ          & (direct):   & 1.232^{\dag}  & 1.204^{\S} & 1.697^{\dag}  & 1.493^{\dag}  \\*[-1pt]
HSE          & (direct):   & 1.172^{\dag}  & 1.132^{\S} & 1.198^{\dag}  & 1.151^{\dag}  \\
HSE          & (indirect): & 2.107^{\ddag} &            & 1.874^{\ddag} & 1.881^{\ddag} \\
Exp.         & (direct):   &               & \multicolumn{1}{c}{~~1.29$^h$; 1.24$^i$}   \\
Calc. GW$^j$ & (direct):   & 1.1^{\dag}    & 1.3^{\S} \\
\end{tabular}
\end{ruledtabular}
\\
Gap nature: 
$^{\dag}{\Gamma\!-\!\Gamma}$, 
$^{\ddag}{\Gamma\!-\!M}$,
$^{\S}{A\!-\!A}$,
$^{\P}{A\!-\!M}$;
experimental data:
$^a$Ref.~\onlinecite{PSSB31-129}, 
$^b$Ref.~\onlinecite{PRB40-3837},
$^c$Ref.~\onlinecite{MaterResBull41-751},
$^d$Ref.~\onlinecite{PhysLettA55-245},
$^e$Ref.~\onlinecite{PRB84-085314},
$^f$Ref.~\onlinecite{JPhChSol74-1240},
$^g$Ref.~\onlinecite{JChemPhys141-084701},
$^h$Ref.~\onlinecite{MaterSciEngB100-263},
$^i$Ref.~\onlinecite{PRB63-125330},
$^j$Ref.~\onlinecite{JPhysChemLett6-3098}. 
In the experimental works\cite{PSSB31-129,PRB40-3837} cited for GaSe,
the distinction between $\gamma$ and $\varepsilon$ phases was not done.
\end{table*} 

\subsection{Band gap character and magnitude}

The calculated band gap values in comparison with available experimental data
are summarized in Table~\ref{tab:gaps}. Compared to PBEsol both HSE and mBJ schemes
augment the gap by $\approx\,1$~eV in GaSe and $\approx\,0.9$~eV in InSe,
setting the values quite close to experimental data. Looking more attentively,
for GaSe the experiment reports an indirect gap (without specifying
its nature) to be slightly smaller than the direct one, for both the $\beta$
and the ($\gamma$ or $\varepsilon$, not clearly identified) polytypes.  
The mBJ calculation yields an astonishing agreement with these subtle details,
assuming the indirect gap between $\Gamma$ and $M$, and the polytype likely
matching the ($\gamma$ or $\varepsilon$) experimental study being the $\gamma$.
In fact, the indirect $\Gamma - M$ gap ($A - M$ in $\gamma$ polytype)
is shorter than the direct one in all four polytypes probed in calculations.
The HSE predictions for the band gap lay close (within several \% to both
the experiments and the mBJ results), however, the direct gap comes out
shorter than the indirect one in all polytypes of GaSe. For InSe,
the band gap seems to be direct according to both the experiment (presumably done on the $\gamma$
phase) and the HSE calculations; the absolute gap value is within \%13 of deviation
from experiment after the HSE calculation and within 7\% after the mBJ calculation.
These observations do not yet necessarily infer that the mBJ is generally more reliable
than the HSE; one should take into account that the ``augmentation'' of the band gap
with respect to the GGA value occurs due to different mechanisms in mBJ and in HSE
formalisms, so that the whole band structure is affected. 
Anyway, mBJ offers a very reasonable accuracy for a calculation cost
much more attractive than that related to HSE.

\section{Conclusion}
\label{sec:conclu}
Summarizing, for GaSe and InSe layered semiconductors we studied the performance,
within the general context of the DFT,
of several prescriptions, now in broad use, for XC potentials, aiming to obtain
reasonable description of the band structures and in particular the band gaps.
In parallel, the performance of these schemes was studied in what concerns
the accurate prediction of the ground-state properties (equilibrium lattice parameters
and -- implicitly -- elastic properties). This latter task was tackled by considering,
in particular, modifications of the DFT total energy aimed at grasping, either 
via additive corrections (the Grimme's, or more flexible schemes), or via realistic
polarization models (MBD scheme), the effect of dispersion interactions.
The tests have been done on layered GaSe and InSe semiconductors,
which have a virtue of being historically well studied, but not so much at the level
of fine differences between their available polytypes. 

We find that among the schemes routinely employed at the level of modern DFT calculations
in view of obtaining reasonable band gaps, namely, mBJ meta-GGA and hybrid HSE functional,
both yield the gap values in good quantitative agreement (within several $\%$)
with experiment and with GW results. We do not find a conclusive evidence in favor
of one of the schemes 
to give systematically better results than the other. Moreover the predictions 
on whether the optical gap is direct or indirect may differ, according to two schemes, 
in view of somehow different
details of band dispersions and a close competition between placing the conduction-band
minimum at the BZ axis ($\Gamma$, $A$) or periphery ($M$, $L$). We point out a noticeable
difference (up to $\sim$ 1~eV) between mBJ and HSE schemes in the placement of semicore states
(Ga $3d$ and $4s$; In $4d$ and $5s$), presumably related to the latter's localization degree.
It would be instructive to compare these predictions with the findings 
from photoemission spectroscopy, of which we could not find any (within the energy range
of interest and sufficient energy resolution) for the systems in question. 

Our other finding concerns the predictions of the equilibrium structures,
and the comparison of corresponding total energies. 
It turns out that whereas the ``conventional'' DFT schemes, e.g., GGA-PBE,
largely overestimate the $a$ and especially the $c$ parameter (hence the interlayer vdW gap),
a considerable improvement (to within $1\%$ of the experimental values) is achieved
by using either PBEsol, or Grimme D2 / D3 / (Becke-Johnson) corrections to PBE.
The discrimination of total energy / volume curves between polytypes,
already quite delicate as assessed in PBEsol calculations, becomes nearly impossible
on inclusion of Grimme corrections (in PBE+D2 calculations), presumably due to 
enhanced sensitivity of such models to interatomic distances between contributing atoms
and not to genuine short-range order and charge density distribution.
However, the calculations done with ``first-principles'' many-body dispersion scheme
do largely recover the discernibility of polytypes.

\section*{Acknowledgments}
The authors thank the PMMS (P{\^o}le Messin de Mod{\'e}lisation et de Simulation) 
and GENCI-/CCRT (Grants x2017-085106 and x2018-085106)
for providing the computational resources.
A.P. acknowledges the support by the R12 Thematic Axis of the Institut Jean Barriol 
(FR2843 CNRS), Universit\'e de Lorraine; 
M.B. and F.EHH -- from the French-Libanese
PHC CEDRE program ``Future Materials''; J.S. -- by the CNRS-Lebanon and
the Lebanese University / Ecole Doctorale des Sciences et de Technologie.


\begin{thebibliography}{100}%
\makeatletter
\providecommand \@ifxundefined [1]{%
 \ifx #1\undefined \expandafter \@firstoftwo
 \else \expandafter \@secondoftwo
\fi
}%
\providecommand \@ifnum [1]{%
 \ifnum #1\expandafter \@firstoftwo
 \else \expandafter \@secondoftwo
\fi
}%
\providecommand \enquote [1]{``#1''}%
\providecommand \bibnamefont  [1]{#1}%
\providecommand \bibfnamefont [1]{#1}%
\providecommand \citenamefont [1]{#1}%
\providecommand\href[0]{\@sanitize\@href}%
\providecommand\@href[1]{\endgroup\@@startlink{#1}\endgroup\@@href}%
\providecommand\@@href[1]{#1\@@endlink}%
\providecommand \@sanitize [0]{\begingroup\catcode`\&12\catcode`\#12\relax}%
\@ifxundefined \pdfoutput {\@firstoftwo}{%
 \@ifnum{\z@=\pdfoutput}{\@firstoftwo}{\@secondoftwo}%
}{%
 \providecommand\@@startlink[1]{\leavevmode}%
 \providecommand\@@endlink[0]{}%
}{%
 \providecommand\@@startlink[1]{%
  \leavevmode
  \pdfstartlink
   attr{/Border[0 0 1 ]/H/I/C[0 1 1]}%
   user{/Subtype/Link/A<</Type/Action/S/URI/URI(#1)>>}%
  \relax
 }%
 \providecommand\@@endlink[0]{\pdfendlink}%
}%
\providecommand \url  [0]{\begingroup\@sanitize \@url }%
\providecommand \@url [1]{\endgroup\@href {#1}{\urlprefix}}%
\providecommand \urlprefix [0]{URL }%
\providecommand \Eprint[0]{\href }%
\@ifxundefined \urlstyle {%
  \providecommand \doi [1]{doi:\discretionary{}{}{}#1}%
}{%
  \providecommand \doi [0]{doi:\discretionary{}{}{}\begingroup
  \urlstyle{rm}\Url }%
}%
\providecommand \doibase [0]{http://dx.doi.org/}%
\providecommand \Doi[1]{\href{\doibase#1}}%
\providecommand \bibAnnote [3]{%
  \BibitemShut{#1}%
  \begin{quotation}\noindent
    \textsc{Key:}\ #2\\\textsc{Annotation:}\ #3%
  \end{quotation}%
}%
\providecommand \bibAnnoteFile [2]{%
  \IfFileExists{#2}{\bibAnnote {#1} {#2} {\input{#2}}}{}%
}%
\providecommand \typeout [0]{\immediate \write \m@ne }%
\providecommand \selectlanguage [0]{\@gobble}%
\providecommand \bibinfo [0]{\@secondoftwo}%
\providecommand \bibfield [0]{\@secondoftwo}%
\providecommand \translation [1]{[#1]}%
\providecommand \BibitemOpen[0]{}%
\providecommand \bibitemStop [0]{}%
\providecommand \bibitemNoStop [0]{.\EOS\space}%
\providecommand \EOS [0]{\spacefactor3000\relax}%
\providecommand \BibitemShut [1]{\csname bibitem#1\endcsname}%
\bibitem{Note1}%
  \BibitemOpen
  \bibinfo {note} {For isolated In-chalcogenide double layers\cite
  {PRB89-205416} and for Ga/In-chalcogenide bi(double)layers,\cite
  {JChemPhys147-114701} a staggered configuration was equally probed in theory
  calculations.}%
  \bibAnnoteFile{Stop}{Note1}%
\bibitem{ACSnano8-1263}%
  \BibitemOpen
  \bibfield{author}{%
  \bibinfo {author} {\bibfnamefont{S.}~\bibnamefont{Lei}}, \bibinfo {author}
  {\bibfnamefont{L.}~\bibnamefont{Ge}}, \bibinfo {author}
  {\bibfnamefont{S.}~\bibnamefont{Najmaei}}, \bibinfo {author}
  {\bibfnamefont{A.}~\bibnamefont{George}}, \bibinfo {author}
  {\bibfnamefont{R.}~\bibnamefont{Kappera}}, \bibinfo {author}
  {\bibfnamefont{J.}~\bibnamefont{Lou}}, \bibinfo {author}
  {\bibfnamefont{M.}~\bibnamefont{Chhowalla}}, \bibinfo {author}
  {\bibfnamefont{H.}~\bibnamefont{Yamaguchi}}, \bibinfo {author}
  {\bibfnamefont{G.}~\bibnamefont{Gupta}}, \bibinfo {author}
  {\bibfnamefont{R.}~\bibnamefont{Vajtai}}, \bibinfo {author}
  {\bibfnamefont{A.~D.}\ \bibnamefont{Mohite}},\ and\ \bibinfo {author}
  {\bibfnamefont{P.~M.}\ \bibnamefont{Ajayan}},\ }%
  \bibfield{journal}{%
  \Doi{10.1021/nn405036u}{\bibinfo {journal} {ACS Nano}}\ }%
  \textbf{\bibinfo {volume} {8}},\ \bibinfo {pages} {1263} (\bibinfo {month}
  {Feb}\ \bibinfo {year} {2014}),\ \url{http://dx.doi.org/10.1021/nn405036u}%
  \bibAnnoteFile{NoStop}{ACSnano8-1263}%
\bibitem{2DMater4-025043}%
  \BibitemOpen
  \bibfield{author}{%
  \bibinfo {author} {\bibfnamefont{N.}~\bibnamefont{Balakrishnan}}, \bibinfo
  {author} {\bibfnamefont{Z.~R.}\ \bibnamefont{Kudrynskyi}}, \bibinfo {author}
  {\bibfnamefont{E.~F.}\ \bibnamefont{Smith}}, \bibinfo {author}
  {\bibfnamefont{M.~W.}\ \bibnamefont{Fay}}, \bibinfo {author}
  {\bibfnamefont{O.}~\bibnamefont{Makarovsky}}, \bibinfo {author}
  {\bibfnamefont{Z.~D.}\ \bibnamefont{Kovalyuk}}, \bibinfo {author}
  {\bibfnamefont{L.}~\bibnamefont{Eaves}}, \bibinfo {author}
  {\bibfnamefont{P.~H.}\ \bibnamefont{Beton}},\ and\ \bibinfo {author}
  {\bibfnamefont{A.}~\bibnamefont{Patan{\`e}}},\ }%
  \bibfield{journal}{%
  \bibinfo {journal} {2D Materials}\ }%
  \textbf{\bibinfo {volume} {4}},\ \bibinfo {pages} {025043} (\bibinfo {month}
  {jun}\ \bibinfo {year} {2017}),\
  \url{http://stacks.iop.org/2053-1583/4/i=2/a=025043}%
  \bibAnnoteFile{NoStop}{2DMater4-025043}%
\bibitem{SciRep4-5497}%
  \BibitemOpen
  \bibfield{author}{%
  \bibinfo {author} {\bibfnamefont{X.}~\bibnamefont{Li}}, \bibinfo {author}
  {\bibfnamefont{M.-W.}\ \bibnamefont{Lin}}, \bibinfo {author}
  {\bibfnamefont{A.~A.}\ \bibnamefont{Puretzky}}, \bibinfo {author}
  {\bibfnamefont{J.~C.}\ \bibnamefont{Idrobo}}, \bibinfo {author}
  {\bibfnamefont{C.}~\bibnamefont{Ma}}, \bibinfo {author}
  {\bibfnamefont{M.}~\bibnamefont{Chi}}, \bibinfo {author}
  {\bibfnamefont{M.}~\bibnamefont{Yoon}}, \bibinfo {author}
  {\bibfnamefont{C.~M.}\ \bibnamefont{Rouleau}}, \bibinfo {author}
  {\bibfnamefont{I.~I.}\ \bibnamefont{Kravchenko}}, \bibinfo {author}
  {\bibfnamefont{D.~B.}\ \bibnamefont{Geohegan}},\ and\ \bibinfo {author}
  {\bibfnamefont{K.}~\bibnamefont{Xiao}},\ }%
  \bibfield{journal}{%
  \Doi{10.1038/srep05497}{\bibinfo {journal} {Sci. Rep.}}\ }%
  \textbf{\bibinfo {volume} {4}},\ \bibinfo {pages} {5497} (\bibinfo {month}
  {Jun}\ \bibinfo {year} {2014}),\ \url{http://dx.doi.org/10.1038/srep05497}%
  \bibAnnoteFile{NoStop}{SciRep4-5497}%
\bibitem{NanoLett13-2777}%
  \BibitemOpen
  \bibfield{author}{%
  \bibinfo {author} {\bibfnamefont{S.}~\bibnamefont{Lei}}, \bibinfo {author}
  {\bibfnamefont{L.}~\bibnamefont{Ge}}, \bibinfo {author}
  {\bibfnamefont{Z.}~\bibnamefont{Liu}}, \bibinfo {author}
  {\bibfnamefont{S.}~\bibnamefont{Najmaei}}, \bibinfo {author}
  {\bibfnamefont{G.}~\bibnamefont{Shi}}, \bibinfo {author}
  {\bibfnamefont{G.}~\bibnamefont{You}}, \bibinfo {author}
  {\bibfnamefont{J.}~\bibnamefont{Lou}}, \bibinfo {author}
  {\bibfnamefont{R.}~\bibnamefont{Vajtai}},\ and\ \bibinfo {author}
  {\bibfnamefont{P.~M.}\ \bibnamefont{Ajayan}},\ }%
  \bibfield{journal}{%
  \Doi{10.1021/nl4010089}{\bibinfo {journal} {Nano Letters}}\ }%
  \textbf{\bibinfo {volume} {13}},\ \bibinfo {pages} {2777} (\bibinfo {month}
  {Jun}\ \bibinfo {year} {2013}),\ \url{http://dx.doi.org/10.1021/nl4010089}%
  \bibAnnoteFile{NoStop}{NanoLett13-2777}%
\bibitem{TSF542-119}%
  \BibitemOpen
  \bibfield{author}{%
  \bibinfo {author} {\bibfnamefont{C.-C.}\ \bibnamefont{Chang}}, \bibinfo
  {author} {\bibfnamefont{J.-X.}\ \bibnamefont{Zeng}}, \bibinfo {author}
  {\bibfnamefont{S.-M.}\ \bibnamefont{Lan}}, \bibinfo {author}
  {\bibfnamefont{W.-Y.}\ \bibnamefont{Uen}}, \bibinfo {author}
  {\bibfnamefont{S.-M.}\ \bibnamefont{Liao}}, \bibinfo {author}
  {\bibfnamefont{T.-N.}\ \bibnamefont{Yang}}, \bibinfo {author}
  {\bibfnamefont{W.-Y.}\ \bibnamefont{Ma}},\ and\ \bibinfo {author}
  {\bibfnamefont{K.-J.}\ \bibnamefont{Chang}},\ }%
  \bibfield{journal}{%
  \Doi{http://dx.doi.org/10.1016/j.tsf.2013.06.087}{\bibinfo {journal} {Thin
  Solid Films}}\ }%
  \textbf{\bibinfo {volume} {542}},\ \bibinfo {pages} {119 } (\bibinfo {month}
  {Sep}\ \bibinfo {year} {2013}),\
  \bibAnnoteFile{NoStop}{TSF542-119}%
\bibitem{PCCP17-10737}%
  \BibitemOpen
  \bibfield{author}{%
  \bibinfo {author} {\bibfnamefont{L.}~\bibnamefont{Ao}}, \bibinfo {author}
  {\bibfnamefont{H.~Y.}\ \bibnamefont{Xiao}}, \bibinfo {author}
  {\bibfnamefont{X.}~\bibnamefont{Xiang}}, \bibinfo {author}
  {\bibfnamefont{S.}~\bibnamefont{Li}}, \bibinfo {author}
  {\bibfnamefont{K.~Z.}\ \bibnamefont{Liu}}, \bibinfo {author}
  {\bibfnamefont{H.}~\bibnamefont{Huang}},\ and\ \bibinfo {author}
  {\bibfnamefont{X.~T.}\ \bibnamefont{Zu}},\ }%
  \bibfield{journal}{%
  \Doi{10.1039/C5CP00397K}{\bibinfo {journal} {Phys. Chem. Chem. Phys.}}\ }%
  \textbf{\bibinfo {volume} {17}},\ \bibinfo {pages} {10737} (\bibinfo {month}
  {apr}\ \bibinfo {year} {2015}),\ \url{http://dx.doi.org/10.1039/C5CP00397K}%
  \bibAnnoteFile{NoStop}{PCCP17-10737}%
\bibitem{AdvMat24-3549}%
  \BibitemOpen
  \bibfield{author}{%
  \bibinfo {author} {\bibfnamefont{D.~J.}\ \bibnamefont{Late}}, \bibinfo
  {author} {\bibfnamefont{B.}~\bibnamefont{Liu}}, \bibinfo {author}
  {\bibfnamefont{J.}~\bibnamefont{Luo}}, \bibinfo {author}
  {\bibfnamefont{A.}~\bibnamefont{Yan}}, \bibinfo {author}
  {\bibfnamefont{H.~S.~S.}\ \bibnamefont{Ramakrishna~Matte}}, \bibinfo {author}
  {\bibfnamefont{M.}~\bibnamefont{Grayson}}, \bibinfo {author}
  {\bibfnamefont{C.~N.~R.}\ \bibnamefont{Rao}},\ and\ \bibinfo {author}
  {\bibfnamefont{V.~P.}\ \bibnamefont{Dravid}},\ }%
  \bibfield{journal}{%
  \Doi{10.1002/adma.201201361}{\bibinfo {journal} {Advanced Materials}}\ }%
  \textbf{\bibinfo {volume} {24}},\ \bibinfo {pages} {3549} (\bibinfo {month}
  {Jul}\ \bibinfo {year} {2012}),\ ISSN \bibinfo {issn} {1521-4095},\
  \url{http://dx.doi.org/10.1002/adma.201201361}%
  \bibAnnoteFile{NoStop}{AdvMat24-3549}%
\bibitem{JACS131-15602}%
  \BibitemOpen
  \bibfield{author}{%
  \bibinfo {author} {\bibfnamefont{J.-J.}\ \bibnamefont{Wang}}, \bibinfo
  {author} {\bibfnamefont{F.-F.}\ \bibnamefont{Cao}}, \bibinfo {author}
  {\bibfnamefont{L.}~\bibnamefont{Jiang}}, \bibinfo {author}
  {\bibfnamefont{Y.-G.}\ \bibnamefont{Guo}}, \bibinfo {author}
  {\bibfnamefont{W.-P.}\ \bibnamefont{Hu}},\ and\ \bibinfo {author}
  {\bibfnamefont{L.-J.}\ \bibnamefont{Wan}},\ }%
  \bibfield{journal}{%
  \Doi{10.1021/ja9072386}{\bibinfo {journal} {Journal of the American Chemical
  Society}}\ }%
  \textbf{\bibinfo {volume} {131}},\ \bibinfo {pages} {15602} (\bibinfo {month}
  {Nov}\ \bibinfo {year} {2009}),\ \url{http://dx.doi.org/10.1021/ja9072386}%
  \bibAnnoteFile{NoStop}{JACS131-15602}%
\bibitem{JChemPhys147-114701}%
  \BibitemOpen
  \bibfield{author}{%
  \bibinfo {author} {\bibfnamefont{T.}~\bibnamefont{Ayadi}}, \bibinfo {author}
  {\bibfnamefont{L.}~\bibnamefont{Debbichi}}, \bibinfo {author}
  {\bibfnamefont{M.}~\bibnamefont{Said}},\ and\ \bibinfo {author}
  {\bibfnamefont{S.}~\bibnamefont{Leb{\`e}gue}},\ }%
  \bibfield{journal}{%
  \Doi{10.1063/1.4997233}{\bibinfo {journal} {The Journal of Chemical
  Physics}}\ }%
  \textbf{\bibinfo {volume} {147}},\ \bibinfo {pages} {114701} (\bibinfo
  {month} {Sep}\ \bibinfo {year} {2017}),\
  \url{https://doi.org/10.1063/1.4997233}%
  \bibAnnoteFile{NoStop}{JChemPhys147-114701}%
\bibitem{PRL102-226401}%
  \BibitemOpen
  \bibfield{author}{%
  \bibinfo {author} {\bibfnamefont{F.}~\bibnamefont{Tran}}\ and\ \bibinfo
  {author} {\bibfnamefont{P.}~\bibnamefont{Blaha}},\ }%
  \bibfield{journal}{%
  \Doi{10.1103/PhysRevLett.102.226401}{\bibinfo {journal} {Phys. Rev. Lett.}}\
  }%
  \textbf{\bibinfo {volume} {102}},\ \bibinfo {pages} {226401} (\bibinfo
  {month} {Jun}\ \bibinfo {year} {2009}),\
  \bibAnnoteFile{NoStop}{PRL102-226401}%
\bibitem{JChemPhys118-8207}%
  \BibitemOpen
  \bibfield{author}{%
  \bibinfo {author} {\bibfnamefont{J.}~\bibnamefont{Heyd}}, \bibinfo {author}
  {\bibfnamefont{G.~E.}\ \bibnamefont{Scuseria}},\ and\ \bibinfo {author}
  {\bibfnamefont{M.}~\bibnamefont{Ernzerhof}},\ }%
  \bibfield{journal}{%
  \Doi{10.1063/1.1564060}{\bibinfo {journal} {The Journal of Chemical
  Physics}}\ }%
  \textbf{\bibinfo {volume} {118}},\ \bibinfo {pages} {8207} (\bibinfo {month}
  {May}\ \bibinfo {year} {2003}),\ \bibinfo {note} {see Erratum
  \cite{JChemPhys124-219906}},\ \url{https://doi.org/10.1063/1.1564060}%
  \bibAnnoteFile{NoStop}{JChemPhys118-8207}%
\bibitem{wien2k}%
  \BibitemOpen
  \bibfield{author}{%
  \bibinfo {author} {\bibfnamefont{P.}~\bibnamefont{Blaha}}, \bibinfo {author}
  {\bibfnamefont{K.}~\bibnamefont{Schwarz}}, \bibinfo {author}
  {\bibfnamefont{G.}~\bibnamefont{Madsen}}, \bibinfo {author}
  {\bibfnamefont{D.}~\bibnamefont{Kvasnicka}},\ and\ \bibinfo {author}
  {\bibfnamefont{J.}~\bibnamefont{Luitz}},\ }%
  \enquote{\bibinfo {title} {{WIEN2k}, an augmented plane wave + local orbitals
  program for calculating crystal properties ({K}arlheinz {S}chwarz, {T}echn.
  {U}niversit\"at {W}ien, {A}ustria, {ISBN 3-9501031-1-2})},}\  (\bibinfo
  {year} {2001}),\ \url{http://www.wien2k.at}%
  \bibAnnoteFile{NoStop}{wien2k}%
\bibitem{vasp}%
  \BibitemOpen
  \bibfield{author}{%
  \bibinfo {author} {\bibfnamefont{G.}~\bibnamefont{Kresse}}, \bibinfo {author}
  {\bibfnamefont{M.}~\bibnamefont{Marsman}},\ and\ \bibinfo {author}
  {\bibfnamefont{J.}~\bibnamefont{Furthm{\"u}ller}},\ }%
  \enquote{\bibinfo {title} {{VASP}, {V}ienna {A}b-initio {S}imulation
  {P}ackage -- the {G}uide. {C}omputational {M}aterials {P}hysics,
  {U}niversit\"at {W}ien, {A}ustria},}\  (\bibinfo {year} {2016}),\
  \url{http://cms.mpi.univie.ac.at/vasp/vasp/vasp.html}%
  \bibAnnoteFile{NoStop}{vasp}%
\bibitem{PSSA31-469}%
  \BibitemOpen
  \bibfield{author}{%
  \bibinfo {author} {\bibfnamefont{A.}~\bibnamefont{Kuhn}}, \bibinfo {author}
  {\bibfnamefont{A.}~\bibnamefont{Chevy}},\ and\ \bibinfo {author}
  {\bibfnamefont{R.}~\bibnamefont{Chevalier}},\ }%
  \bibfield{journal}{%
  \Doi{10.1002/pssa.2210310216}{\bibinfo {journal} {physica status solidi
  (a)}}\ }%
  \textbf{\bibinfo {volume} {31}},\ \bibinfo {pages} {469} (\bibinfo {month}
  {Oct}\ \bibinfo {year} {1975}),\ ISSN \bibinfo {issn} {1521-396X},\
  \url{http://dx.doi.org/10.1002/pssa.2210310216}%
  \bibAnnoteFile{NoStop}{PSSA31-469}%
\bibitem{ComRenAcadSciC279-33}%
  \BibitemOpen
  \bibfield{author}{%
  \bibinfo {author} {\bibfnamefont{A.}~\bibnamefont{Likforman}}\ and\ \bibinfo
  {author} {\bibfnamefont{M.}~\bibnamefont{Guittard}},\ }%
  \bibfield{journal}{%
  \bibinfo {journal} {Comptes Rendus des S\'eances de l'Acad\'emie des Sciences
  -- Paris, S\'erie C}\ }%
  \textbf{\bibinfo {volume} {279}},\ \bibinfo {pages} {33} (\bibinfo {month}
  {Jul}\ \bibinfo {year} {1974}),\
  \bibAnnoteFile{NoStop}{ComRenAcadSciC279-33}%
\bibitem{PRB15-3200}%
  \BibitemOpen
  \bibfield{author}{%
  \bibinfo {author} {\bibfnamefont{P.~K.}\ \bibnamefont{Larsen}}, \bibinfo
  {author} {\bibfnamefont{S.}~\bibnamefont{Chiang}},\ and\ \bibinfo {author}
  {\bibfnamefont{N.~V.}\ \bibnamefont{Smith}},\ }%
  \bibfield{journal}{%
  \Doi{10.1103/PhysRevB.15.3200}{\bibinfo {journal} {Phys. Rev. B}}\ }%
  \textbf{\bibinfo {volume} {15}},\ \bibinfo {pages} {3200} (\bibinfo {month}
  {Mar}\ \bibinfo {year} {1977}),\
  \url{http://link.aps.org/doi/10.1103/PhysRevB.15.3200}%
  \bibAnnoteFile{NoStop}{PRB15-3200}%
\bibitem{JPCM11-4303}%
  \BibitemOpen
  \bibfield{author}{%
  \bibinfo {author} {\bibfnamefont{A.}~\bibnamefont{Amokrane}}, \bibinfo
  {author} {\bibfnamefont{F.}~\bibnamefont{Proix}}, \bibinfo {author}
  {\bibfnamefont{S.}~\bibnamefont{El~{M}onkad}}, \bibinfo {author}
  {\bibfnamefont{A.}~\bibnamefont{Cricenti}}, \bibinfo {author}
  {\bibfnamefont{C.}~\bibnamefont{Barchesi}}, \bibinfo {author}
  {\bibfnamefont{M.}~\bibnamefont{Eddrief}}, \bibinfo {author}
  {\bibfnamefont{K.}~\bibnamefont{Amimer}},\ and\ \bibinfo {author}
  {\bibfnamefont{C.~A.}\ \bibnamefont{S{\'e}benne}},\ }%
  \bibfield{journal}{%
  \Doi{10.1088/0953-8984/11/22/302}{\bibinfo {journal} {Journal of Physics:
  Condensed Matter}}\ }%
  \textbf{\bibinfo {volume} {11}},\ \bibinfo {pages} {4303} (\bibinfo {month}
  {Jun}\ \bibinfo {year} {1999}),\
  \bibAnnoteFile{NoStop}{JPCM11-4303}%
\bibitem{SSC22-685}%
  \BibitemOpen
  \bibfield{author}{%
  \bibinfo {author} {\bibfnamefont{P.}~\bibnamefont{Thiry}}, \bibinfo {author}
  {\bibfnamefont{Y.}~\bibnamefont{Petroff}}, \bibinfo {author}
  {\bibfnamefont{R.}~\bibnamefont{Pinchaux}}, \bibinfo {author}
  {\bibfnamefont{C.}~\bibnamefont{Guillot}}, \bibinfo {author}
  {\bibfnamefont{Y.}~\bibnamefont{Ballu}}, \bibinfo {author}
  {\bibfnamefont{J.}~\bibnamefont{Lecante}}, \bibinfo {author}
  {\bibfnamefont{J.}~\bibnamefont{Paign{\'e}}},\ and\ \bibinfo {author}
  {\bibfnamefont{F.}~\bibnamefont{Levy}},\ }%
  \bibfield{journal}{%
  \Doi{http://dx.doi.org/10.1016/0038-1098(77)90250-2}{\bibinfo {journal}
  {Solid State Communications}}\ }%
  \textbf{\bibinfo {volume} {22}},\ \bibinfo {pages} {685 } (\bibinfo {month}
  {Jun}\ \bibinfo {year} {1977}),\ ISSN \bibinfo {issn} {0038-1098},\
  \bibAnnoteFile{NoStop}{SSC22-685}%
\bibitem{PRB68-125304}%
  \BibitemOpen
  \bibfield{author}{%
  \bibinfo {author} {\bibfnamefont{L.}~\bibnamefont{Plucinski}}, \bibinfo
  {author} {\bibfnamefont{R.~L.}\ \bibnamefont{Johnson}}, \bibinfo {author}
  {\bibfnamefont{B.~J.}\ \bibnamefont{Kowalski}}, \bibinfo {author}
  {\bibfnamefont{K.}~\bibnamefont{Kopalko}}, \bibinfo {author}
  {\bibfnamefont{B.~A.}\ \bibnamefont{Orlowski}}, \bibinfo {author}
  {\bibfnamefont{Z.~D.}\ \bibnamefont{Kovalyuk}},\ and\ \bibinfo {author}
  {\bibfnamefont{G.~V.}\ \bibnamefont{Lashkarev}},\ }%
  \bibfield{journal}{%
  \Doi{10.1103/PhysRevB.68.125304}{\bibinfo {journal} {Phys. Rev. B}}\ }%
  \textbf{\bibinfo {volume} {68}},\ \bibinfo {pages} {125304} (\bibinfo {month}
  {Sep}\ \bibinfo {year} {2003}),\
  \url{http://link.aps.org/doi/10.1103/PhysRevB.68.125304}%
  \bibAnnoteFile{NoStop}{PRB68-125304}%
\bibitem{PRB49-11093}%
  \BibitemOpen
  \bibfield{author}{%
  \bibinfo {author} {\bibfnamefont{R.}~\bibnamefont{Sporken}}, \bibinfo
  {author} {\bibfnamefont{R.}~\bibnamefont{Hafsi}}, \bibinfo {author}
  {\bibfnamefont{F.}~\bibnamefont{Coletti}}, \bibinfo {author}
  {\bibfnamefont{J.~M.}\ \bibnamefont{Debever}}, \bibinfo {author}
  {\bibfnamefont{P.~A.}\ \bibnamefont{Thiry}},\ and\ \bibinfo {author}
  {\bibfnamefont{A.}~\bibnamefont{Chevy}},\ }%
  \bibfield{journal}{%
  \Doi{10.1103/PhysRevB.49.11093}{\bibinfo {journal} {Phys. Rev. B}}\ }%
  \textbf{\bibinfo {volume} {49}},\ \bibinfo {pages} {11093} (\bibinfo {month}
  {Apr}\ \bibinfo {year} {1994}),\
  \url{https://link.aps.org/doi/10.1103/PhysRevB.49.11093}%
  \bibAnnoteFile{NoStop}{PRB49-11093}%
\bibitem{PRB13-3534}%
  \BibitemOpen
  \bibfield{author}{%
  \bibinfo {author} {\bibfnamefont{M.}~\bibnamefont{Schl\"uter}}, \bibinfo
  {author} {\bibfnamefont{J.}~\bibnamefont{Camassel}}, \bibinfo {author}
  {\bibfnamefont{S.}~\bibnamefont{Kohn}}, \bibinfo {author}
  {\bibfnamefont{J.~P.}\ \bibnamefont{Voitchovsky}}, \bibinfo {author}
  {\bibfnamefont{Y.~R.}\ \bibnamefont{Shen}},\ and\ \bibinfo {author}
  {\bibfnamefont{M.~L.}\ \bibnamefont{Cohen}},\ }%
  \bibfield{journal}{%
  \Doi{10.1103/PhysRevB.13.3534}{\bibinfo {journal} {Phys. Rev. B}}\ }%
  \textbf{\bibinfo {volume} {13}},\ \bibinfo {pages} {3534} (\bibinfo {month}
  {Apr}\ \bibinfo {year} {1976}),\
  \url{http://link.aps.org/doi/10.1103/PhysRevB.13.3534}%
  \bibAnnoteFile{NoStop}{PRB13-3534}%
\bibitem{PRB14-424}%
  \BibitemOpen
  \bibfield{author}{%
  \bibinfo {author} {\bibfnamefont{M.}~\bibnamefont{Schl{\"u}ter}}\ and\
  \bibinfo {author} {\bibfnamefont{M.~L.}\ \bibnamefont{Cohen}},\ }%
  \bibfield{journal}{%
  \Doi{10.1103/PhysRevB.14.424}{\bibinfo {journal} {Phys. Rev. B}}\ }%
  \textbf{\bibinfo {volume} {14}},\ \bibinfo {pages} {424} (\bibinfo {month}
  {Jul}\ \bibinfo {year} {1976}),\
  \url{http://link.aps.org/doi/10.1103/PhysRevB.14.424}%
  \bibAnnoteFile{NoStop}{PRB14-424}%
\bibitem{JPhysC10-1211}%
  \BibitemOpen
  \bibfield{author}{%
  \bibinfo {author} {\bibfnamefont{J.~V.}\ \bibnamefont{{McC}anny}}\ and\
  \bibinfo {author} {\bibfnamefont{R.~B.}\ \bibnamefont{Murray}},\ }%
  \bibfield{journal}{%
  \Doi{10.1088/0022-3719/10/8/022}{\bibinfo {journal} {Journal of Physics C:
  Solid State Physics}}\ }%
  \textbf{\bibinfo {volume} {10}},\ \bibinfo {pages} {1211} (\bibinfo {month}
  {Apr}\ \bibinfo {year} {1977}),\
  \url{http://stacks.iop.org/0022-3719/10/i=8/a=022}%
  \bibAnnoteFile{NoStop}{JPhysC10-1211}%
\bibitem{SSC27-1449}%
  \BibitemOpen
  \bibfield{author}{%
  \bibinfo {author} {\bibfnamefont{Y.}~\bibnamefont{Depeursinge}}, \bibinfo
  {author} {\bibfnamefont{E.}~\bibnamefont{Doni}}, \bibinfo {author}
  {\bibfnamefont{R.}~\bibnamefont{Girlanda}}, \bibinfo {author}
  {\bibfnamefont{A.}~\bibnamefont{Baldereschi}},\ and\ \bibinfo {author}
  {\bibfnamefont{K.~.}\ \bibnamefont{Maschke}},\ }%
  \bibfield{journal}{%
  \Doi{http://dx.doi.org/10.1016/0038-1098(78)91593-4}{\bibinfo {journal}
  {Solid State Communications}}\ }%
  \textbf{\bibinfo {volume} {27}},\ \bibinfo {pages} {1449 } (\bibinfo {month}
  {Sep}\ \bibinfo {year} {1978}),\
  \bibAnnoteFile{NoStop}{SSC27-1449}%
\bibitem{NuovoCimB51-154}%
  \BibitemOpen
  \bibfield{author}{%
  \bibinfo {author} {\bibfnamefont{E.}~\bibnamefont{Doni}}, \bibinfo {author}
  {\bibfnamefont{R.}~\bibnamefont{Girlanda}}, \bibinfo {author}
  {\bibfnamefont{V.}~\bibnamefont{Grasso}}, \bibinfo {author}
  {\bibfnamefont{A.}~\bibnamefont{Balzarotti}},\ and\ \bibinfo {author}
  {\bibfnamefont{M.}~\bibnamefont{Piacentini}},\ }%
  \bibfield{journal}{%
  \Doi{10.1007/BF02743704}{\bibinfo {journal} {Il Nuovo Cimento B}}\ }%
  \textbf{\bibinfo {volume} {51}},\ \bibinfo {pages} {154} (\bibinfo {month}
  {May}\ \bibinfo {year} {1979}),\ \url{http://dx.doi.org/10.1007/BF02743704}%
\bibitem{JPhysC12-1625}%
  \BibitemOpen
  \bibfield{author}{%
  \bibinfo {author} {\bibfnamefont{S.}~\bibnamefont{Nagel}}, \bibinfo {author}
  {\bibfnamefont{A.}~\bibnamefont{Baldereschi}},\ and\ \bibinfo {author}
  {\bibfnamefont{K.}~\bibnamefont{Maschke}},\ }%
  \bibfield{journal}{%
  \bibinfo {journal} {Journal of Physics C: Solid State Physics}\ }%
  \textbf{\bibinfo {volume} {12}},\ \bibinfo {pages} {1625} (\bibinfo {month}
  {may}\ \bibinfo {year} {1979}),\
  \url{http://stacks.iop.org/0022-3719/12/i=9/a=006}%
  \bibAnnoteFile{NoStop}{JPhysC12-1625}%
\bibitem{JPhysC12-4777}%
  \BibitemOpen
  \bibfield{author}{%
  \bibinfo {author} {\bibfnamefont{J.}~\bibnamefont{Robertson}},\ }%
  \bibfield{journal}{%
  \Doi{10.1088/0022-3719/12/22/019}{\bibinfo {journal} {Journal of Physics C:
  Solid State Physics}}\ }%
  \textbf{\bibinfo {volume} {12}},\ \bibinfo {pages} {4777} (\bibinfo {month}
  {Nov}\ \bibinfo {year} {1979}),\
  \bibAnnoteFile{NoStop}{JPhysC12-4777}%
\bibitem{PhysicaBC105-324}%
  \BibitemOpen
  \bibfield{author}{%
  \bibinfo {author} {\bibfnamefont{Y.}~\bibnamefont{Depeursinge}}\ and\
  \bibinfo {author} {\bibfnamefont{A.}~\bibnamefont{Baldereschi}},\ }%
  \bibfield{journal}{%
  \Doi{http://dx.doi.org/10.1016/0378-4363(81)90268-0}{\bibinfo {journal}
  {Physica B+C}}\ }%
  \textbf{\bibinfo {volume} {105}},\ \bibinfo {pages} {324 } (\bibinfo {month}
  {May}\ \bibinfo {year} {1981}),\ ISSN \bibinfo {issn} {0378-4363},\
  \bibAnnoteFile{NoStop}{PhysicaBC105-324}%
\bibitem{PRB48-14135}%
  \BibitemOpen
  \bibfield{author}{%
  \bibinfo {author} {\bibfnamefont{P.}~\bibnamefont{Gomes {da}~Costa}},
  \bibinfo {author} {\bibfnamefont{R.~G.}\ \bibnamefont{Dandrea}}, \bibinfo
  {author} {\bibfnamefont{R.~F.}\ \bibnamefont{Wallis}},\ and\ \bibinfo
  {author} {\bibfnamefont{M.}~\bibnamefont{Balkanski}},\ }%
  \bibfield{journal}{%
  \Doi{10.1103/PhysRevB.48.14135}{\bibinfo {journal} {Phys. Rev. B}}\ }%
  \textbf{\bibinfo {volume} {48}},\ \bibinfo {pages} {14135} (\bibinfo {month}
  {Nov}\ \bibinfo {year} {1993}),\
  \url{http://link.aps.org/doi/10.1103/PhysRevB.48.14135}%
  \bibAnnoteFile{NoStop}{PRB48-14135}%
\bibitem{PRB57-3726}%
  \BibitemOpen
  \bibfield{author}{%
  \bibinfo {author} {\bibfnamefont{C.}~\bibnamefont{Adler}}, \bibinfo {author}
  {\bibfnamefont{R.}~\bibnamefont{Honke}}, \bibinfo {author}
  {\bibfnamefont{P.}~\bibnamefont{Pavone}},\ and\ \bibinfo {author}
  {\bibfnamefont{U.}~\bibnamefont{Schr\"oder}},\ }%
  \bibfield{journal}{%
  \Doi{10.1103/PhysRevB.57.3726}{\bibinfo {journal} {Phys. Rev. B}}\ }%
  \textbf{\bibinfo {volume} {57}},\ \bibinfo {pages} {3726} (\bibinfo {month}
  {Feb}\ \bibinfo {year} {1998}),\
  \url{http://link.aps.org/doi/10.1103/PhysRevB.57.3726}%
  \bibAnnoteFile{NoStop}{PRB57-3726}%
\bibitem{JPCM11-6715}%
  \BibitemOpen
  \bibfield{author}{%
  \bibinfo {author} {\bibfnamefont{S.-W.}\ \bibnamefont{Yu}}, \bibinfo {author}
  {\bibfnamefont{T.}~\bibnamefont{Lischke}}, \bibinfo {author}
  {\bibfnamefont{N.}~\bibnamefont{M{\"u}ller}}, \bibinfo {author}
  {\bibfnamefont{U.}~\bibnamefont{Heinzmann}}, \bibinfo {author}
  {\bibfnamefont{C.}~\bibnamefont{Pettenkofer}}, \bibinfo {author}
  {\bibfnamefont{A.}~\bibnamefont{Klein}}, \bibinfo {author}
  {\bibfnamefont{P.}~\bibnamefont{Blaha}},\ and\ \bibinfo {author}
  {\bibfnamefont{J.}~\bibnamefont{Braun}},\ }%
  \bibfield{journal}{%
  \Doi{10.1088/0953-8984/11/35/310}{\bibinfo {journal} {Journal of Physics:
  Condensed Matter}}\ }%
  \textbf{\bibinfo {volume} {11}},\ \bibinfo {pages} {6715} (\bibinfo {month}
  {Sep}\ \bibinfo {year} {1999}),\
  \bibAnnoteFile{NoStop}{JPCM11-6715}%
\bibitem{ChinPhysLett23-1876}%
  \BibitemOpen
  \bibfield{author}{%
  \bibinfo {author} {\bibfnamefont{D.-W.}\ \bibnamefont{Zhang}}, \bibinfo
  {author} {\bibfnamefont{F.-T.}\ \bibnamefont{Jin}},\ and\ \bibinfo {author}
  {\bibfnamefont{J.-M.}\ \bibnamefont{Yuan}},\ }%
  \bibfield{journal}{%
  \bibinfo {journal} {Chinese Physics Letters}\ }%
  \textbf{\bibinfo {volume} {23}},\ \bibinfo {pages} {1876} (\bibinfo {month}
  {Jul}\ \bibinfo {year} {2006}),\
  \url{http://stacks.iop.org/0256-307X/23/i=7/a=061}%
  \bibAnnoteFile{NoStop}{ChinPhysLett23-1876}%
\bibitem{PRB84-085314}%
  \BibitemOpen
  \bibfield{author}{%
  \bibinfo {author} {\bibfnamefont{D.~V.}\ \bibnamefont{Rybkovskiy}}, \bibinfo
  {author} {\bibfnamefont{N.~R.}\ \bibnamefont{Arutyunyan}}, \bibinfo {author}
  {\bibfnamefont{A.~S.}\ \bibnamefont{Orekhov}}, \bibinfo {author}
  {\bibfnamefont{I.~A.}\ \bibnamefont{Gromchenko}}, \bibinfo {author}
  {\bibfnamefont{I.~V.}\ \bibnamefont{Vorobiev}}, \bibinfo {author}
  {\bibfnamefont{A.~V.}\ \bibnamefont{Osadchy}}, \bibinfo {author}
  {\bibfnamefont{E.~Y.}\ \bibnamefont{Salaev}}, \bibinfo {author}
  {\bibfnamefont{T.~K.}\ \bibnamefont{Baykara}}, \bibinfo {author}
  {\bibfnamefont{K.~R.}\ \bibnamefont{Allakhverdiev}},\ and\ \bibinfo {author}
  {\bibfnamefont{E.~D.}\ \bibnamefont{Obraztsova}},\ }%
  \bibfield{journal}{%
  \Doi{10.1103/PhysRevB.84.085314}{\bibinfo {journal} {Phys. Rev. B}}\ }%
  \textbf{\bibinfo {volume} {84}},\ \bibinfo {pages} {085314} (\bibinfo {month}
  {Aug}\ \bibinfo {year} {2011}),\
  \url{http://link.aps.org/doi/10.1103/PhysRevB.84.085314}%
  \bibAnnoteFile{NoStop}{PRB84-085314}%
\bibitem{CompMatSci67-73}%
  \BibitemOpen
  \bibfield{author}{%
  \bibinfo {author} {\bibfnamefont{L.}~\bibnamefont{Ghalouci}}, \bibinfo
  {author} {\bibfnamefont{B.}~\bibnamefont{Benbahi}}, \bibinfo {author}
  {\bibfnamefont{S.}~\bibnamefont{Hiadsi}}, \bibinfo {author}
  {\bibfnamefont{B.}~\bibnamefont{Abidri}}, \bibinfo {author}
  {\bibfnamefont{G.}~\bibnamefont{Vergoten}},\ and\ \bibinfo {author}
  {\bibfnamefont{F.}~\bibnamefont{Ghalouci}},\ }%
  \bibfield{journal}{%
  \Doi{http://dx.doi.org/10.1016/j.commatsci.2012.08.034}{\bibinfo {journal}
  {Computational Materials Science}}\ }%
  \textbf{\bibinfo {volume} {67}},\ \bibinfo {pages} {10} (\bibinfo {month}
  {feb}\ \bibinfo {year} {2013}),\
  \bibAnnoteFile{NoStop}{CompMatSci67-73}%
\bibitem{CompMatSci124-62}%
  \BibitemOpen
  \bibfield{author}{%
  \bibinfo {author} {\bibfnamefont{L.}~\bibnamefont{Ghalouci}}, \bibinfo
  {author} {\bibfnamefont{F.}~\bibnamefont{Taibi}}, \bibinfo {author}
  {\bibfnamefont{F.}~\bibnamefont{Ghalouci}},\ and\ \bibinfo {author}
  {\bibfnamefont{M.~O.}\ \bibnamefont{Bensaid}},\ }%
  \bibfield{journal}{%
  \Doi{https://doi.org/10.1016/j.commatsci.2016.07.013}{\bibinfo {journal}
  {Computational Materials Science}}\ }%
  \textbf{\bibinfo {volume} {124}},\ \bibinfo {pages} {62} (\bibinfo {month}
  {Nov}\ \bibinfo {year} {2016}),\ ISSN \bibinfo {issn} {0927-0256},\
  \bibAnnoteFile{NoStop}{CompMatSci124-62}%
\bibitem{PCCP15-7098}%
  \BibitemOpen
  \bibfield{author}{%
  \bibinfo {author} {\bibfnamefont{Y.}~\bibnamefont{Ma}}, \bibinfo {author}
  {\bibfnamefont{Y.}~\bibnamefont{Dai}}, \bibinfo {author}
  {\bibfnamefont{M.}~\bibnamefont{Guo}}, \bibinfo {author}
  {\bibfnamefont{L.}~\bibnamefont{Yu}},\ and\ \bibinfo {author}
  {\bibfnamefont{B.}~\bibnamefont{Huang}},\ }%
  \bibfield{journal}{%
  \Doi{10.1039/c3cp50233c}{\bibinfo {journal} {Phys. Chem. Chem. Phys.}}\ }%
  \textbf{\bibinfo {volume} {15}},\ \bibinfo {pages} {7098} (\bibinfo {month}
  {may}\ \bibinfo {year} {2013}),\ \url{http://dx.doi.org/10.1039/c3cp50233c}%
  \bibAnnoteFile{NoStop}{PCCP15-7098}%
\bibitem{EurPhysJB86-350}%
  \BibitemOpen
  \bibfield{author}{%
  \bibinfo {author} {\bibfnamefont{D.}~\bibnamefont{Olgu{\'\i}n}}, \bibinfo
  {author} {\bibfnamefont{A.}~\bibnamefont{Rubio-Ponce}},\ and\ \bibinfo
  {author} {\bibfnamefont{A.}~\bibnamefont{Cantarero}},\ }%
  \bibfield{journal}{%
  \Doi{10.1140/epjb/e2013-40141-1}{\bibinfo {journal} {The European Physical
  Journal B}}\ }%
  \textbf{\bibinfo {volume} {86}},\ \bibinfo {pages} {350} (\bibinfo {month}
  {Aug}\ \bibinfo {year} {2013}),\ ISSN \bibinfo {issn} {1434-6028},\
  \url{http://dx.doi.org/10.1140/epjb/e2013-40141-1}%
  \bibAnnoteFile{NoStop}{EurPhysJB86-350}%
\bibitem{JPhChSol70-344}%
  \BibitemOpen
  \bibfield{author}{%
  \bibinfo {author} {\bibfnamefont{Z.}~\bibnamefont{Rak}}, \bibinfo {author}
  {\bibfnamefont{S.~D.}\ \bibnamefont{Mahanti}}, \bibinfo {author}
  {\bibfnamefont{K.~C.}\ \bibnamefont{Mandal}},\ and\ \bibinfo {author}
  {\bibfnamefont{N.~C.}\ \bibnamefont{Fernelius}},\ }%
  \bibfield{journal}{%
  \Doi{http://dx.doi.org/10.1016/j.jpcs.2008.10.022}{\bibinfo {journal}
  {Journal of Physics and Chemistry of Solids}}\ }%
  \textbf{\bibinfo {volume} {70}},\ \bibinfo {pages} {344} (\bibinfo {month}
  {feb}\ \bibinfo {year} {2009}),\ ISSN \bibinfo {issn} {0022-3697},\
  \bibAnnoteFile{NoStop}{JPhChSol70-344}%
\bibitem{PhysicaB436-188}%
  \BibitemOpen
  \bibfield{author}{%
  \bibinfo {author} {\bibfnamefont{S.-R.}\ \bibnamefont{Zhang}}, \bibinfo
  {author} {\bibfnamefont{S.-F.}\ \bibnamefont{Zhu}}, \bibinfo {author}
  {\bibfnamefont{B.-J.}\ \bibnamefont{Zhao}}, \bibinfo {author}
  {\bibfnamefont{L.-H.}\ \bibnamefont{Xie}},\ and\ \bibinfo {author}
  {\bibfnamefont{K.}~\bibnamefont{e~Hui~Song}},\ }%
  \bibfield{journal}{%
  \Doi{http://dx.doi.org/10.1016/j.physb.2013.12.014}{\bibinfo {journal}
  {Physica B: Condensed Matter}}\ }%
  \textbf{\bibinfo {volume} {436}},\ \bibinfo {pages} {188 } (\bibinfo {month}
  {Mar}\ \bibinfo {year} {2014}),\ ISSN \bibinfo {issn} {0921-4526},\
  \bibAnnoteFile{NoStop}{PhysicaB436-188}%
\bibitem{PRB90-235302}%
  \BibitemOpen
  \bibfield{author}{%
  \bibinfo {author} {\bibfnamefont{D.~V.}\ \bibnamefont{Rybkovskiy}}, \bibinfo
  {author} {\bibfnamefont{A.~V.}\ \bibnamefont{Osadchy}},\ and\ \bibinfo
  {author} {\bibfnamefont{E.~D.}\ \bibnamefont{Obraztsova}},\ }%
  \bibfield{journal}{%
  \Doi{10.1103/PhysRevB.90.235302}{\bibinfo {journal} {Phys. Rev. B}}\ }%
  \textbf{\bibinfo {volume} {90}},\ \bibinfo {pages} {235302} (\bibinfo {month}
  {Dec}\ \bibinfo {year} {2014}),\
  \url{http://link.aps.org/doi/10.1103/PhysRevB.90.235302}%
  \bibAnnoteFile{NoStop}{PRB90-235302}%
\bibitem{JPhysChemLett6-3098}%
  \BibitemOpen
  \bibfield{author}{%
  \bibinfo {author} {\bibfnamefont{L.}~\bibnamefont{Debbichi}}, \bibinfo
  {author} {\bibfnamefont{O.}~\bibnamefont{Eriksson}},\ and\ \bibinfo {author}
  {\bibfnamefont{S.}~\bibnamefont{Leb{\`e}gue}},\ }%
  \bibfield{journal}{%
  \Doi{10.1021/acs.jpclett.5b01356}{\bibinfo {journal} {The Journal of Physical
  Chemistry Letters}}\ }%
  \textbf{\bibinfo {volume} {6}},\ \bibinfo {pages} {3098} (\bibinfo {month}
  {aug}\ \bibinfo {year} {2015}),\
  \url{http://dx.doi.org/10.1021/acs.jpclett.5b01356}%
  \bibAnnoteFile{NoStop}{JPhysChemLett6-3098}%
\bibitem{PRB63-125330}%
  \BibitemOpen
  \bibfield{author}{%
  \bibinfo {author} {\bibfnamefont{F.~J.}\ \bibnamefont{Manj{\'o}n}}, \bibinfo
  {author} {\bibfnamefont{D.}~\bibnamefont{Errandonea}}, \bibinfo {author}
  {\bibfnamefont{A.}~\bibnamefont{Segura}}, \bibinfo {author}
  {\bibfnamefont{V.}~\bibnamefont{Mu{\~n}oz}}, \bibinfo {author}
  {\bibfnamefont{G.}~\bibnamefont{Tob{\'{\i}}as}}, \bibinfo {author}
  {\bibfnamefont{P.}~\bibnamefont{Ordej{\'o}n}},\ and\ \bibinfo {author}
  {\bibfnamefont{E.}~\bibnamefont{Canadell}},\ }%
  \bibfield{journal}{%
  \Doi{10.1103/PhysRevB.63.125330}{\bibinfo {journal} {Phys. Rev. B}}\ }%
  \textbf{\bibinfo {volume} {63}},\ \bibinfo {pages} {125330} (\bibinfo {month}
  {Mar}\ \bibinfo {year} {2001}),\
  \url{http://link.aps.org/doi/10.1103/PhysRevB.63.125330}%
  \bibAnnoteFile{NoStop}{PRB63-125330}%
\bibitem{PRB66-085210}%
  \BibitemOpen
  \bibfield{author}{%
  \bibinfo {author} {\bibfnamefont{G.}~\bibnamefont{Ferlat}}, \bibinfo {author}
  {\bibfnamefont{H.}~\bibnamefont{Xu}}, \bibinfo {author}
  {\bibfnamefont{V.}~\bibnamefont{Timoshevskii}},\ and\ \bibinfo {author}
  {\bibfnamefont{X.}~\bibnamefont{Blase}},\ }%
  \bibfield{journal}{%
  \Doi{10.1103/PhysRevB.66.085210}{\bibinfo {journal} {Phys. Rev. B}}\ }%
  \textbf{\bibinfo {volume} {66}},\ \bibinfo {pages} {085210} (\bibinfo {month}
  {Aug}\ \bibinfo {year} {2002}),\
  \url{http://link.aps.org/doi/10.1103/PhysRevB.66.085210}%
  \bibAnnoteFile{NoStop}{PRB66-085210}%
\bibitem{PRB71-125206}%
  \BibitemOpen
  \bibfield{author}{%
  \bibinfo {author} {\bibfnamefont{D.}~\bibnamefont{Errandonea}}, \bibinfo
  {author} {\bibfnamefont{A.}~\bibnamefont{Segura}}, \bibinfo {author}
  {\bibfnamefont{F.~J.}\ \bibnamefont{Manj{\'o}n}}, \bibinfo {author}
  {\bibfnamefont{A.}~\bibnamefont{Chevy}}, \bibinfo {author}
  {\bibfnamefont{E.}~\bibnamefont{Machado}}, \bibinfo {author}
  {\bibfnamefont{G.}~\bibnamefont{Tobias}}, \bibinfo {author}
  {\bibfnamefont{P.}~\bibnamefont{Ordej{\'o}n}},\ and\ \bibinfo {author}
  {\bibfnamefont{E.}~\bibnamefont{Canadell}},\ }%
  \bibfield{journal}{%
  \Doi{10.1103/PhysRevB.71.125206}{\bibinfo {journal} {Phys. Rev. B}}\ }%
  \textbf{\bibinfo {volume} {71}},\ \bibinfo {pages} {125206} (\bibinfo {month}
  {Mar}\ \bibinfo {year} {2005}),\
  \url{http://link.aps.org/doi/10.1103/PhysRevB.71.125206}%
  \bibAnnoteFile{NoStop}{PRB71-125206}%
\bibitem{PSSB244-244}%
  \BibitemOpen
  \bibfield{author}{%
  \bibinfo {author} {\bibfnamefont{U.}~\bibnamefont{Schwarz}}, \bibinfo
  {author} {\bibfnamefont{D.}~\bibnamefont{Olguin}}, \bibinfo {author}
  {\bibfnamefont{A.}~\bibnamefont{Cantarero}}, \bibinfo {author}
  {\bibfnamefont{M.}~\bibnamefont{Hanfland}},\ and\ \bibinfo {author}
  {\bibfnamefont{K.}~\bibnamefont{Syassen}},\ }%
  \bibfield{journal}{%
  \Doi{10.1002/pssb.200672551}{\bibinfo {journal} {physica status solidi (b)}}\
  }%
  \textbf{\bibinfo {volume} {244}},\ \bibinfo {pages} {244} (\bibinfo {month}
  {jan}\ \bibinfo {year} {2007}),\ ISSN \bibinfo {issn} {1521-3951},\
  \url{http://dx.doi.org/10.1002/pssb.200672551}%
  \bibAnnoteFile{NoStop}{PSSB244-244}%
\bibitem{PRB77-045208}%
  \BibitemOpen
  \bibfield{author}{%
  \bibinfo {author} {\bibfnamefont{D.}~\bibnamefont{Errandonea}}, \bibinfo
  {author} {\bibfnamefont{D.}~\bibnamefont{Mart{\'{\i}}nez-Garc{\'{i}}a}},
  \bibinfo {author} {\bibfnamefont{A.}~\bibnamefont{Segura}}, \bibinfo {author}
  {\bibfnamefont{J.}~\bibnamefont{Haines}}, \bibinfo {author}
  {\bibfnamefont{E.}~\bibnamefont{Machado-Charry}}, \bibinfo {author}
  {\bibfnamefont{E.}~\bibnamefont{Canadell}}, \bibinfo {author}
  {\bibfnamefont{J.~C.}\ \bibnamefont{Chervin}},\ and\ \bibinfo {author}
  {\bibfnamefont{A.}~\bibnamefont{Chevy}},\ }%
  \bibfield{journal}{%
  \Doi{10.1103/PhysRevB.77.045208}{\bibinfo {journal} {Phys. Rev. B}}\ }%
  \textbf{\bibinfo {volume} {77}},\ \bibinfo {pages} {045208} (\bibinfo {month}
  {Jan}\ \bibinfo {year} {2008}),\
  \url{http://link.aps.org/doi/10.1103/PhysRevB.77.045208}%
  \bibAnnoteFile{NoStop}{PRB77-045208}%
\bibitem{PRB70-125201}%
  \BibitemOpen
  \bibfield{author}{%
  \bibinfo {author} {\bibfnamefont{F.~J.}\ \bibnamefont{Manj{\'o}n}}, \bibinfo
  {author} {\bibfnamefont{A.}~\bibnamefont{Segura}}, \bibinfo {author}
  {\bibfnamefont{V.}~\bibnamefont{Mu{\~n}oz-Sanjos{\'e}}}, \bibinfo {author}
  {\bibfnamefont{G.}~\bibnamefont{Tob{\'{\i}}as}}, \bibinfo {author}
  {\bibfnamefont{P.}~\bibnamefont{Ordej{\'o}n}},\ and\ \bibinfo {author}
  {\bibfnamefont{E.}~\bibnamefont{Canadell}},\ }%
  \bibfield{journal}{%
  \Doi{10.1103/PhysRevB.70.125201}{\bibinfo {journal} {Phys. Rev. B}}\ }%
  \textbf{\bibinfo {volume} {70}},\ \bibinfo {pages} {125201} (\bibinfo {month}
  {Sep}\ \bibinfo {year} {2004}),\
  \url{http://link.aps.org/doi/10.1103/PhysRevB.70.125201}%
  \bibAnnoteFile{NoStop}{PRB70-125201}%
\bibitem{Semicon44-1158}%
  \BibitemOpen
  \bibfield{author}{%
  \bibinfo {author} {\bibfnamefont{V.~N.}\ \bibnamefont{Brudnyi}}, \bibinfo
  {author} {\bibfnamefont{A.~V.}\ \bibnamefont{Kosobutsky}},\ and\ \bibinfo
  {author} {\bibfnamefont{S.~Y.}\ \bibnamefont{Sarkisov}},\ }%
  \bibfield{journal}{%
  \Doi{10.1134/S1063782610090095}{\bibinfo {journal} {Semiconductors}}\ }%
  \textbf{\bibinfo {volume} {44}},\ \bibinfo {pages} {1158} (\bibinfo {month}
  {sep}\ \bibinfo {year} {2010}),\ ISSN \bibinfo {issn} {1063-7826},\ \bibinfo
  {note} {published in: Fizika i Tekhnika Poluprovodnikov, Vol. 44, No. 9
  (2010), pp. 1194 -- 1202},\
  \url{http://dx.doi.org/10.1134/S1063782610090095}%
  \bibAnnoteFile{NoStop}{Semicon44-1158}%
\bibitem{PRL108-266805}%
  \BibitemOpen
  \bibfield{author}{%
  \bibinfo {author} {\bibfnamefont{Z.}~\bibnamefont{Zhu}}, \bibinfo {author}
  {\bibfnamefont{Y.}~\bibnamefont{Cheng}},\ and\ \bibinfo {author}
  {\bibfnamefont{U.}~\bibnamefont{Schwingenschl\"ogl}},\ }%
  \bibfield{journal}{%
  \Doi{10.1103/PhysRevLett.108.266805}{\bibinfo {journal} {Phys. Rev. Lett.}}\
  }%
  \textbf{\bibinfo {volume} {108}},\ \bibinfo {pages} {266805} (\bibinfo
  {month} {Jun}\ \bibinfo {year} {2012}),\
  \bibAnnoteFile{NoStop}{PRL108-266805}%
\bibitem{JPhChSol74-1240}%
  \BibitemOpen
  \bibfield{author}{%
  \bibinfo {author} {\bibfnamefont{A.~V.}\ \bibnamefont{Kosobutsky}}, \bibinfo
  {author} {\bibfnamefont{S.~Y.}\ \bibnamefont{Sarkisov}},\ and\ \bibinfo
  {author} {\bibfnamefont{V.~N.}\ \bibnamefont{Brudnyi}},\ }%
  \bibfield{journal}{%
  \Doi{http://dx.doi.org/10.1016/j.jpcs.2013.03.025}{\bibinfo {journal}
  {Journal of Physics and Chemistry of Solids}}\ }%
  \textbf{\bibinfo {volume} {74}},\ \bibinfo {pages} {1240 } (\bibinfo {month}
  {Sep}\ \bibinfo {year} {2013}),\ ISSN \bibinfo {issn} {0022-3697},\
  \bibAnnoteFile{NoStop}{JPhChSol74-1240}%
\bibitem{PhysSolState46-179}%
  \BibitemOpen
  \bibfield{author}{%
  \bibinfo {author} {\bibfnamefont{K.~Z.}\ \bibnamefont{Rushchanski{\u{\i}}}},\
  }%
  \bibfield{journal}{%
  \Doi{10.1134/1.1641949}{\bibinfo {journal} {Physics of the Solid State}}\ }%
  \textbf{\bibinfo {volume} {46}},\ \bibinfo {pages} {179} (\bibinfo {month}
  {Jan}\ \bibinfo {year} {2004}),\ ISSN \bibinfo {issn} {1063-7834},\
  \url{http://dx.doi.org/10.1134/1.1641949}%
  \bibAnnoteFile{NoStop}{PhysSolState46-179}%
\bibitem{JChemPhys141-084701}%
  \BibitemOpen
  \bibfield{author}{%
  \bibinfo {author} {\bibfnamefont{W.}~\bibnamefont{An}}, \bibinfo {author}
  {\bibfnamefont{F.}~\bibnamefont{Wu}}, \bibinfo {author}
  {\bibfnamefont{H.}~\bibnamefont{Jiang}}, \bibinfo {author}
  {\bibfnamefont{G.-S.}\ \bibnamefont{Tian}},\ and\ \bibinfo {author}
  {\bibfnamefont{X.-Z.}\ \bibnamefont{Li}},\ }%
  \bibfield{journal}{%
  \Doi{http://dx.doi.org/10.1063/1.4893346}{\bibinfo {journal} {The Journal of
  Chemical Physics}}\ }%
  \textbf{\bibinfo {volume} {141}},\ \bibinfo {eid} {084701} (\bibinfo {month}
  {Aug}\ \bibinfo {year} {2014}),\
  \bibAnnoteFile{NoStop}{JChemPhys141-084701}%
\bibitem{PRB47-558}%
  \BibitemOpen
  \bibfield{author}{%
  \bibinfo {author} {\bibfnamefont{G.}~\bibnamefont{Kresse}}\ and\ \bibinfo
  {author} {\bibfnamefont{J.}~\bibnamefont{Hafner}},\ }%
  \bibfield{journal}{%
  \Doi{10.1103/PhysRevB.47.558}{\bibinfo {journal} {Phys. Rev. B}}\ }%
  \textbf{\bibinfo {volume} {47}},\ \bibinfo {pages} {558} (\bibinfo {month}
  {Jan}\ \bibinfo {year} {1993}),\
  \url{http://link.aps.org/doi/10.1103/PhysRevB.47.558}%
  \bibAnnoteFile{NoStop}{PRB47-558}%
\bibitem{PRB54-11169}%
  \BibitemOpen
  \bibfield{author}{%
  \bibinfo {author} {\bibfnamefont{G.}~\bibnamefont{Kresse}}\ and\ \bibinfo
  {author} {\bibfnamefont{J.}~\bibnamefont{Furthm{\"u}ller}},\ }%
  \bibfield{journal}{%
  \Doi{10.1103/PhysRevB.54.11169}{\bibinfo {journal} {Phys. Rev. B}}\ }%
  \textbf{\bibinfo {volume} {54}},\ \bibinfo {pages} {11169} (\bibinfo {month}
  {Oct}\ \bibinfo {year} {1996}),\
  \url{http://link.aps.org/doi/10.1103/PhysRevB.54.11169}%
  \bibAnnoteFile{NoStop}{PRB54-11169}%
\bibitem{PRB50-17953}%
  \BibitemOpen
  \bibfield{author}{%
  \bibinfo {author} {\bibfnamefont{P.~E.}\ \bibnamefont{Bl{\"o}chl}},\ }%
  \bibfield{journal}{%
  \Doi{10.1103/PhysRevB.50.17953}{\bibinfo {journal} {Phys. Rev. B}}\ }%
  \textbf{\bibinfo {volume} {50}},\ \bibinfo {pages} {17953} (\bibinfo {month}
  {Dec}\ \bibinfo {year} {1994}),\
  \url{http://link.aps.org/doi/10.1103/PhysRevB.50.17953}%
  \bibAnnoteFile{NoStop}{PRB50-17953}%
\bibitem{PRB59-1758}%
  \BibitemOpen
  \bibfield{author}{%
  \bibinfo {author} {\bibfnamefont{G.}~\bibnamefont{Kresse}}\ and\ \bibinfo
  {author} {\bibfnamefont{D.}~\bibnamefont{Joubert}},\ }%
  \bibfield{journal}{%
  \Doi{10.1103/PhysRevB.59.1758}{\bibinfo {journal} {Phys. Rev. B}}\ }%
  \textbf{\bibinfo {volume} {59}},\ \bibinfo {pages} {1758} (\bibinfo {month}
  {Jan}\ \bibinfo {year} {1999}),\
  \url{http://link.aps.org/doi/10.1103/PhysRevB.59.1758}%
  \bibAnnoteFile{NoStop}{PRB59-1758}%
\bibitem{JCompChem29-2044}%
  \BibitemOpen
  \bibfield{author}{%
  \bibinfo {author} {\bibfnamefont{J.}~\bibnamefont{Hafner}},\ }%
  \bibfield{journal}{%
  \Doi{10.1002/jcc.21057}{\bibinfo {journal} {Journal of Computational
  Chemistry}}\ }%
  \textbf{\bibinfo {volume} {29}},\ \bibinfo {pages} {2044} (\bibinfo {month}
  {Oct}\ \bibinfo {year} {2008}),\ ISSN \bibinfo {issn} {1096-987X},\
  \url{http://dx.doi.org/10.1002/jcc.21057}%
  \bibAnnoteFile{NoStop}{JCompChem29-2044}%
\bibitem{SSC9-1763}%
  \BibitemOpen
  \bibfield{author}{%
  \bibinfo {author} {\bibfnamefont{O.}~\bibnamefont{Jepsen}}\ and\ \bibinfo
  {author} {\bibfnamefont{O.~K.}\ \bibnamefont{Andersen}},\ }%
  \bibfield{journal}{%
  \Doi{http://dx.doi.org/10.1016/0038-1098(71)90313-9}{\bibinfo {journal}
  {Solid State Communications}}\ }%
  \textbf{\bibinfo {volume} {9}},\ \bibinfo {pages} {1763} (\bibinfo {year}
  {1971}),\ ISSN \bibinfo {issn} {0038-1098},\ 
  \bibAnnoteFile{NoStop}{SSC9-1763}%
\bibitem{PRB49-16223}%
  \BibitemOpen
  \bibfield{author}{%
  \bibinfo {author} {\bibfnamefont{P.~E.}\ \bibnamefont{Bl{\"o}chl}}, \bibinfo
  {author} {\bibfnamefont{O.}~\bibnamefont{Jepsen}},\ and\ \bibinfo {author}
  {\bibfnamefont{O.~K.}\ \bibnamefont{Andersen}},\ }%
  \bibfield{journal}{%
  \Doi{10.1103/PhysRevB.49.16223}{\bibinfo {journal} {Phys. Rev. B}}\ }%
  \textbf{\bibinfo {volume} {49}},\ \bibinfo {pages} {16223} (\bibinfo {month}
  {Jun}\ \bibinfo {year} {1994}),\
  \url{http://link.aps.org/doi/10.1103/PhysRevB.49.16223}%
  \bibAnnoteFile{NoStop}{PRB49-16223}%
\bibitem{PRB13-5188}%
  \BibitemOpen
  \bibfield{author}{%
  \bibinfo {author} {\bibfnamefont{H.~J.}\ \bibnamefont{Monkhorst}}\ and\
  \bibinfo {author} {\bibfnamefont{J.~D.}\ \bibnamefont{Pack}},\ }%
  \bibfield{journal}{%
  \Doi{10.1103/PhysRevB.13.5188}{\bibinfo {journal} {Phys. Rev. B}}\ }%
  \textbf{\bibinfo {volume} {13}},\ \bibinfo {pages} {5188} (\bibinfo {month}
  {Jun}\ \bibinfo {year} {1976}),\
  \url{http://link.aps.org/doi/10.1103/PhysRevB.13.5188}%
  \bibAnnoteFile{NoStop}{PRB13-5188}%
\bibitem{PSSB254-1700120}%
  \BibitemOpen
  \bibfield{author}{%
  \bibinfo {author} {\bibfnamefont{J.}~\bibnamefont{Srour}}, \bibinfo {author}
  {\bibfnamefont{A.}~\bibnamefont{Postnikov}}, \bibinfo {author}
  {\bibfnamefont{M.}~\bibnamefont{Badawi}},\ and\ \bibinfo {author}
  {\bibfnamefont{F.}~\bibnamefont{El~Haj~Hassan}},\ }%
  \bibfield{journal}{%
  \Doi{10.1002/pssb.201700120}{\bibinfo {journal} {physica status solidi (b)}}\
  }%
  \textbf{\bibinfo {volume} {254}},\ \bibinfo {pages} {1700120} (\bibinfo
  {month} {jun}\ \bibinfo {year} {2017}),\ ISSN \bibinfo {issn} {1521-3951},\
  \url{http://dx.doi.org/10.1002/pssb.201700120}%
  \bibAnnoteFile{NoStop}{PSSB254-1700120}%
\bibitem{PRL77-3865}%
  \BibitemOpen
  \bibfield{author}{%
  \bibinfo {author} {\bibfnamefont{J.~P.}\ \bibnamefont{Perdew}}, \bibinfo
  {author} {\bibfnamefont{K.}~\bibnamefont{Burke}},\ and\ \bibinfo {author}
  {\bibfnamefont{M.}~\bibnamefont{Ernzerhof}},\ }%
  \bibfield{journal}{%
  \Doi{10.1103/PhysRevLett.77.3865}{\bibinfo {journal} {Phys. Rev. Lett.}}\ }%
  \textbf{\bibinfo {volume} {77}},\ \bibinfo {pages} {3865} (\bibinfo {month}
  {Oct}\ \bibinfo {year} {1996}),\ \bibinfo {note} {see Erratum
  \cite{PRL78-1396}, Comment \cite{PRL80-890} and Reply \cite{PRL80-891}},\
  \url{http://dx.doi.org/10.1103/PhysRevLett.77.3865}%
  \bibAnnoteFile{NoStop}{PRL77-3865}%
\bibitem{PRL78-1396}%
  \BibitemOpen
  \bibfield{author}{%
  \bibinfo {author} {\bibfnamefont{J.~P.}\ \bibnamefont{Perdew}}, \bibinfo
  {author} {\bibfnamefont{K.}~\bibnamefont{Burke}},\ and\ \bibinfo {author}
  {\bibfnamefont{M.}~\bibnamefont{Ernzerhof}},\ }%
  \bibfield{journal}{%
  \bibinfo {journal} {Phys. Rev. Lett.}\ }%
  \textbf{\bibinfo {volume} {78}},\ \bibinfo {pages} {1396} (\bibinfo {month}
  {Feb}\ \bibinfo {year} {1997}),\
  \url{http://dx.doi.org/10.1103/PhysRevLett.78.1396}%
  \bibAnnoteFile{NoStop}{PRL78-1396}%
\bibitem{PRL100-136406}%
  \BibitemOpen
  \bibfield{author}{%
  \bibinfo {author} {\bibfnamefont{J.~P.}\ \bibnamefont{Perdew}}, \bibinfo
  {author} {\bibfnamefont{A.}~\bibnamefont{Ruzsinszky}}, \bibinfo {author}
  {\bibfnamefont{G.~I.}\ \bibnamefont{Csonka}}, \bibinfo {author}
  {\bibfnamefont{O.~A.}\ \bibnamefont{Vydrov}}, \bibinfo {author}
  {\bibfnamefont{G.~E.}\ \bibnamefont{Scuseria}}, \bibinfo {author}
  {\bibfnamefont{L.~A.}\ \bibnamefont{Constantin}}, \bibinfo {author}
  {\bibfnamefont{X.}~\bibnamefont{Zhou}},\ and\ \bibinfo {author}
  {\bibfnamefont{K.}~\bibnamefont{Burke}},\ }%
  \bibfield{journal}{%
  \Doi{10.1103/PhysRevLett.100.136406}{\bibinfo {journal} {Phys. Rev. Lett.}}\
  }%
  \textbf{\bibinfo {volume} {100}},\ \bibinfo {pages} {136406} (\bibinfo
  {month} {Apr}\ \bibinfo {year} {2008}),\ \bibinfo {note} {the basic Ref. to
  PBEsol; see also \cite{PRL102-039902,PRL101-239701,PRL101-239702}},\
  \bibAnnoteFile{NoStop}{PRL100-136406}%
\bibitem{Juliana_thesis}%
  \BibitemOpen
  \bibfield{author}{%
  \bibinfo {author} {\bibfnamefont{J.~Y.}\ \bibnamefont{Srour}},\ }%
  \emph{\bibinfo {title} {Electronic structure and competition of phases in
  {Cu-(In,Ga)-Se}, {Ga-Se} and {In-Se} semiconductors: first-principles
  calculations based on different exchange-correlation potentials}},\ Ph.D.
  thesis,\ \bibinfo {school} {Universit\'e de Lorraine} (\bibinfo {year}
  {2016}),\ \url{http://theses.fr/2016LORR0238}%
  \bibAnnoteFile{NoStop}{Juliana_thesis}%
\bibitem{PRL92-246401}%
  \BibitemOpen
  \bibfield{author}{%
  \bibinfo {author} {\bibfnamefont{M.}~\bibnamefont{Dion}}, \bibinfo {author}
  {\bibfnamefont{H.}~\bibnamefont{Rydberg}}, \bibinfo {author}
  {\bibfnamefont{E.}~\bibnamefont{Schr\"oder}}, \bibinfo {author}
  {\bibfnamefont{D.~C.}\ \bibnamefont{Langreth}},\ and\ \bibinfo {author}
  {\bibfnamefont{B.~I.}\ \bibnamefont{Lundqvist}},\ }%
  \bibfield{journal}{%
  \Doi{10.1103/PhysRevLett.92.246401}{\bibinfo {journal} {Phys. Rev. Lett.}}\
  }%
  \textbf{\bibinfo {volume} {92}},\ \bibinfo {pages} {246401} (\bibinfo {month}
  {Jun}\ \bibinfo {year} {2004}),\
  \bibAnnoteFile{NoStop}{PRL92-246401}%
\bibitem{PRB82-081101}%
  \BibitemOpen
  \bibfield{author}{%
  \bibinfo {author} {\bibfnamefont{K.}~\bibnamefont{Lee}}, \bibinfo {author}
  {\bibfnamefont{E.~D.}\ \bibnamefont{Murray}}, \bibinfo {author}
  {\bibfnamefont{L.}~\bibnamefont{Kong}}, \bibinfo {author}
  {\bibfnamefont{B.~I.}\ \bibnamefont{Lundqvist}},\ and\ \bibinfo {author}
  {\bibfnamefont{D.~C.}\ \bibnamefont{Langreth}},\ }%
  \bibfield{journal}{%
  \Doi{10.1103/PhysRevB.82.081101}{\bibinfo {journal} {Phys. Rev. B}}\ }%
  \textbf{\bibinfo {volume} {82}},\ \bibinfo {pages} {081101} (\bibinfo {month}
  {Aug}\ \bibinfo {year} {2010}),\
  \url{https://link.aps.org/doi/10.1103/PhysRevB.82.081101}%
  \bibAnnoteFile{NoStop}{PRB82-081101}%
\bibitem{PRB83-195131}%
  \BibitemOpen
  \bibfield{author}{%
  \bibinfo {author} {\bibfnamefont{J.}~\bibnamefont{Klime{\v{s}}}}, \bibinfo
  {author} {\bibfnamefont{D.~R.}\ \bibnamefont{Bowler}},\ and\ \bibinfo
  {author} {\bibfnamefont{A.}~\bibnamefont{Michaelides}},\ }%
  \bibfield{journal}{%
  \Doi{10.1103/PhysRevB.83.195131}{\bibinfo {journal} {Phys. Rev. B}}\ }%
  \textbf{\bibinfo {volume} {83}},\ \bibinfo {pages} {195131} (\bibinfo {month}
  {May}\ \bibinfo {year} {2011}),\
  \url{https://link.aps.org/doi/10.1103/PhysRevB.83.195131}%
  \bibAnnoteFile{NoStop}{PRB83-195131}%
\bibitem{JPhysChemA114-11814}%
  \BibitemOpen
  \bibfield{author}{%
  \bibinfo {author} {\bibfnamefont{T.}~\bibnamefont{Bu{\v{c}}ko}}, \bibinfo
  {author} {\bibfnamefont{J.}~\bibnamefont{Hafner}}, \bibinfo {author}
  {\bibfnamefont{S.}~\bibnamefont{Leb{\`e}gue}},\ and\ \bibinfo {author}
  {\bibfnamefont{J.}~\bibnamefont{{\'A}ngy{\'a}n}},\ }%
  \bibfield{journal}{%
  \Doi{10.1021/jp106469x}{\bibinfo {journal} {The Journal of Physical Chemistry
  A}}\ }%
  \textbf{\bibinfo {volume} {114}},\ \bibinfo {pages} {11814} (\bibinfo {month}
  {Nov}\ \bibinfo {year} {2010}),\ \url{http://dx.doi.org/10.1021/jp106469x}%
  \bibAnnoteFile{NoStop}{JPhysChemA114-11814}%
\bibitem{JChemTheoComp9-4293}%
  \BibitemOpen
  \bibfield{author}{%
  \bibinfo {author} {\bibfnamefont{T.}~\bibnamefont{Bu{\v{c}}ko}}, \bibinfo
  {author} {\bibfnamefont{S.}~\bibnamefont{Leb{\`e}gue}}, \bibinfo {author}
  {\bibfnamefont{J.}~\bibnamefont{Hafner}},\ and\ \bibinfo {author}
  {\bibfnamefont{J.~G.}\ \bibnamefont{{\'A}ngy{\'a}n}},\ }%
  \bibfield{journal}{%
  \Doi{10.1021/ct400694h}{\bibinfo {journal} {Journal of Chemical Theory and
  Computation}}\ }%
  \textbf{\bibinfo {volume} {9}},\ \bibinfo {pages} {4293} (\bibinfo {month}
  {Oct}\ \bibinfo {year} {2013}),\ \url{http://dx.doi.org/10.1021/ct400694h}%
  \bibAnnoteFile{NoStop}{JChemTheoComp9-4293}%
\bibitem{JChemPhys141-034114}%
  \BibitemOpen
  \bibfield{author}{%
  \bibinfo {author} {\bibfnamefont{T.}~\bibnamefont{Bu{\v{c}}ko}}, \bibinfo
  {author} {\bibfnamefont{S.}~\bibnamefont{Leb{\`e}gue}}, \bibinfo {author}
  {\bibfnamefont{J.~G.}\ \bibnamefont{{\'A}ngy{\'a}n}},\ and\ \bibinfo {author}
  {\bibfnamefont{J.}~\bibnamefont{Hafner}},\ }%
  \bibfield{journal}{%
  \Doi{10.1063/1.4890003}{\bibinfo {journal} {The Journal of Chemical
  Physics}}\ }%
  \textbf{\bibinfo {volume} {141}},\ \bibinfo {pages} {034114} (\bibinfo
  {month} {Jul}\ \bibinfo {year} {2014}),\
  \url{https://doi.org/10.1063/1.4890003}%
  \bibAnnoteFile{NoStop}{JChemPhys141-034114}%
\bibitem{JPCM28-045201}%
  \BibitemOpen
  \bibfield{author}{%
  \bibinfo {author} {\bibfnamefont{T.}~\bibnamefont{Bu{\v{c}}ko}}, \bibinfo
  {author} {\bibfnamefont{S.}~\bibnamefont{Leb{\`e}gue}}, \bibinfo {author}
  {\bibfnamefont{T.}~\bibnamefont{Gould}},\ and\ \bibinfo {author}
  {\bibfnamefont{J.~G.}\ \bibnamefont{{\'{A}}ngy{\'a}n}},\ }%
  \bibfield{journal}{%
  \bibinfo {journal} {Journal of Physics: Condensed Matter}\ }%
  \textbf{\bibinfo {volume} {28}},\ \bibinfo {pages} {045201} (\bibinfo {month}
  {Feb}\ \bibinfo {year} {2016}),\
  \url{http://stacks.iop.org/0953-8984/28/i=4/a=045201}%
  \bibAnnoteFile{NoStop}{JPCM28-045201}%
\bibitem{JChemTheoComp12-5920}%
  \BibitemOpen
  \bibfield{author}{%
  \bibinfo {author} {\bibfnamefont{T.}~\bibnamefont{Gould}}, \bibinfo {author}
  {\bibfnamefont{S.}~\bibnamefont{Leb{\`e}gue}}, \bibinfo {author}
  {\bibfnamefont{J.~G.}\ \bibnamefont{{\'A}ngy{\'a}n}},\ and\ \bibinfo {author}
  {\bibfnamefont{T.}~\bibnamefont{Bu{\v{c}}ko}},\ }%
  \bibfield{journal}{%
  \Doi{10.1021/acs.jctc.6b00925}{\bibinfo {journal} {Journal of Chemical Theory
  and Computation}}\ }%
  \textbf{\bibinfo {volume} {12}},\ \bibinfo {pages} {5920} (\bibinfo {month}
  {Dec}\ \bibinfo {year} {2016}),\ \bibinfo {note} {pMID: 27951673},\
  \url{http://dx.doi.org/10.1021/acs.jctc.6b00925}%
  \bibAnnoteFile{NoStop}{JChemTheoComp12-5920}%
\bibitem{JCompChem27-1787}%
  \BibitemOpen
  \bibfield{author}{%
  \bibinfo {author} {\bibfnamefont{S.}~\bibnamefont{Grimme}},\ }%
  \bibfield{journal}{%
  \Doi{10.1002/jcc.20495}{\bibinfo {journal} {Journal of Computational
  Chemistry}}\ }%
  \textbf{\bibinfo {volume} {27}},\ \bibinfo {pages} {1787} (\bibinfo {month}
  {Nov}\ \bibinfo {year} {2006}),\ ISSN \bibinfo {issn} {1096-987X},\
  \url{http://dx.doi.org/10.1002/jcc.20495}%
  \bibAnnoteFile{NoStop}{JCompChem27-1787}%
\bibitem{JChemPhys132-154104}%
  \BibitemOpen
  \bibfield{author}{%
  \bibinfo {author} {\bibfnamefont{S.}~\bibnamefont{Grimme}}, \bibinfo {author}
  {\bibfnamefont{J.}~\bibnamefont{Antony}}, \bibinfo {author}
  {\bibfnamefont{S.}~\bibnamefont{Ehrlich}},\ and\ \bibinfo {author}
  {\bibfnamefont{H.}~\bibnamefont{Krieg}},\ }%
  \bibfield{journal}{%
  \Doi{10.1063/1.3382344}{\bibinfo {journal} {The Journal of Chemical
  Physics}}\ }%
  \textbf{\bibinfo {volume} {132}},\ \bibinfo {pages} {154104} (\bibinfo
  {month} {Apr}\ \bibinfo {year} {2010}),\
  \url{https://doi.org/10.1063/1.3382344}%
  \bibAnnoteFile{NoStop}{JChemPhys132-154104}%
\bibitem{JCompChem32-1456}%
  \BibitemOpen
  \bibfield{author}{%
  \bibinfo {author} {\bibfnamefont{S.}~\bibnamefont{Grimme}}, \bibinfo {author}
  {\bibfnamefont{S.}~\bibnamefont{Ehrlich}},\ and\ \bibinfo {author}
  {\bibfnamefont{L.}~\bibnamefont{Goerigk}},\ }%
  \bibfield{journal}{%
  \Doi{10.1002/jcc.21759}{\bibinfo {journal} {Journal of Computational
  Chemistry}}\ }%
  \textbf{\bibinfo {volume} {32}},\ \bibinfo {pages} {1456} (\bibinfo {month}
  {May}\ \bibinfo {year} {2011}),\ ISSN \bibinfo {issn} {1096-987X},\
  \url{http://dx.doi.org/10.1002/jcc.21759}%
  \bibAnnoteFile{NoStop}{JCompChem32-1456}%
\bibitem{JChemPhys123-024101}%
  \BibitemOpen
  \bibfield{author}{%
  \bibinfo {author} {\bibfnamefont{E.~R.}\ \bibnamefont{Johnson}}\ and\
  \bibinfo {author} {\bibfnamefont{A.~D.}\ \bibnamefont{Becke}},\ }%
  \bibfield{journal}{%
  \Doi{10.1063/1.1949201}{\bibinfo {journal} {The Journal of Chemical
  Physics}}\ }%
  \textbf{\bibinfo {volume} {123}},\ \bibinfo {pages} {024101} (\bibinfo
  {month} {Jul}\ \bibinfo {year} {2005}),\
  \url{https://doi.org/10.1063/1.1949201}%
  \bibAnnoteFile{NoStop}{JChemPhys123-024101}%
\bibitem{PRL102-073005}%
  \BibitemOpen
  \bibfield{author}{%
  \bibinfo {author} {\bibfnamefont{A.}~\bibnamefont{Tkatchenko}}\ and\ \bibinfo
  {author} {\bibfnamefont{M.}~\bibnamefont{Scheffler}},\ }%
  \bibfield{journal}{%
  \Doi{10.1103/PhysRevLett.102.073005}{\bibinfo {journal} {Phys. Rev. Lett.}}\
  }%
  \textbf{\bibinfo {volume} {102}},\ \bibinfo {pages} {073005} (\bibinfo
  {month} {Feb}\ \bibinfo {year} {2009}),\
  \bibAnnoteFile{NoStop}{PRL102-073005}%
\bibitem{TheoChimActa44-129}%
  \BibitemOpen
  \bibfield{author}{%
  \bibinfo {author} {\bibfnamefont{F.~L.}\ \bibnamefont{Hirshfeld}},\ }%
  \bibfield{journal}{%
  \Doi{10.1007/BF00549096}{\bibinfo {journal} {Theoretica Chimica Acta}}\ }%
  \textbf{\bibinfo {volume} {44}},\ \bibinfo {pages} {129} (\bibinfo {month}
  {Jun}\ \bibinfo {year} {1977}),\ \url{https://doi.org/10.1007/BF00549096}%
  \bibAnnoteFile{NoStop}{TheoChimActa44-129}%
\bibitem{JChemPhys126-144111}%
  \BibitemOpen
  \bibfield{author}{%
  \bibinfo {author} {\bibfnamefont{P.}~\bibnamefont{Bultinck}}, \bibinfo
  {author} {\bibfnamefont{C.}~\bibnamefont{Van~{A}lsenoy}}, \bibinfo {author}
  {\bibfnamefont{P.~W.}\ \bibnamefont{Ayers}},\ and\ \bibinfo {author}
  {\bibfnamefont{R.}~\bibnamefont{Carb{\'o}-Dorca}},\ }%
  \bibfield{journal}{%
  \Doi{10.1063/1.2715563}{\bibinfo {journal} {The Journal of Chemical
  Physics}}\ }%
  \textbf{\bibinfo {volume} {126}},\ \bibinfo {pages} {144111} (\bibinfo
  {month} {Apr}\ \bibinfo {year} {2007}),\
  \url{https://doi.org/10.1063/1.2715563}%
  \bibAnnoteFile{NoStop}{JChemPhys126-144111}%
\bibitem{PRL108-236402}%
  \BibitemOpen
  \bibfield{author}{%
  \bibinfo {author} {\bibfnamefont{A.}~\bibnamefont{Tkatchenko}}, \bibinfo
  {author} {\bibfnamefont{R.~A.}\ \bibnamefont{{{DiS}tasio, Jr.}}}, \bibinfo
  {author} {\bibfnamefont{R.}~\bibnamefont{Car}},\ and\ \bibinfo {author}
  {\bibfnamefont{M.}~\bibnamefont{Scheffler}},\ }%
  \bibfield{journal}{%
  \Doi{10.1103/PhysRevLett.108.236402}{\bibinfo {journal} {Phys. Rev. Lett.}}\
  }%
  \textbf{\bibinfo {volume} {108}},\ \bibinfo {pages} {236402} (\bibinfo
  {month} {Jun}\ \bibinfo {year} {2012}),\
  \bibAnnoteFile{NoStop}{PRL108-236402}%
\bibitem{JChemPhys140-18A508}%
  \BibitemOpen
  \bibfield{author}{%
  \bibinfo {author} {\bibfnamefont{A.}~\bibnamefont{Ambrosetti}}, \bibinfo
  {author} {\bibfnamefont{A.~M.}\ \bibnamefont{Reilly}}, \bibinfo {author}
  {\bibfnamefont{R.~A.}\ \bibnamefont{{{DiS}tasio, Jr.}}},\ and\ \bibinfo
  {author} {\bibfnamefont{A.}~\bibnamefont{Tkatchenko}},\ }%
  \bibfield{journal}{%
  \Doi{10.1063/1.4865104}{\bibinfo {journal} {The Journal of Chemical
  Physics}}\ }%
  \textbf{\bibinfo {volume} {140}},\ \bibinfo {pages} {18A508} (\bibinfo
  {month} {May}\ \bibinfo {year} {2014}),\
  \url{https://doi.org/10.1063/1.4865104}%
  \bibAnnoteFile{NoStop}{JChemPhys140-18A508}%
\bibitem{PRL103-096102}%
  \BibitemOpen
  \bibfield{author}{%
  \bibinfo {author} {\bibfnamefont{G.}~\bibnamefont{Rom\'an-P\'erez}}\ and\
  \bibinfo {author} {\bibfnamefont{J.~M.}\ \bibnamefont{Soler}},\ }%
  \bibfield{journal}{%
  \Doi{10.1103/PhysRevLett.103.096102}{\bibinfo {journal} {Phys. Rev. Lett.}}\
  }%
  \textbf{\bibinfo {volume} {103}},\ \bibinfo {pages} {096102} (\bibinfo
  {month} {Aug}\ \bibinfo {year} {2009}),\
  \bibAnnoteFile{NoStop}{PRL103-096102}%
\bibitem{PRB96-054103}%
  \BibitemOpen
  \bibfield{author}{%
  \bibinfo {author} {\bibfnamefont{F.}~\bibnamefont{Tran}}, \bibinfo {author}
  {\bibfnamefont{J.}~\bibnamefont{Stelzl}}, \bibinfo {author}
  {\bibfnamefont{D.}~\bibnamefont{Koller}}, \bibinfo {author}
  {\bibfnamefont{T.}~\bibnamefont{Ruh}},\ and\ \bibinfo {author}
  {\bibfnamefont{P.}~\bibnamefont{Blaha}},\ }%
  \bibfield{journal}{%
  \Doi{10.1103/PhysRevB.96.054103}{\bibinfo {journal} {Phys. Rev. B}}\ }%
  \textbf{\bibinfo {volume} {96}},\ \bibinfo {pages} {054103} (\bibinfo {month}
  {Aug}\ \bibinfo {year} {2017}),\
  \bibAnnoteFile{NoStop}{PRB96-054103}%
\bibitem{JChemPhys124-219906}%
  \BibitemOpen
  \bibfield{author}{%
  \bibinfo {author} {\bibfnamefont{J.}~\bibnamefont{Heyd}}, \bibinfo {author}
  {\bibfnamefont{G.~E.}\ \bibnamefont{Scuseria}},\ and\ \bibinfo {author}
  {\bibfnamefont{M.}~\bibnamefont{Ernzerhof}},\ }%
  \bibfield{journal}{%
  \Doi{10.1063/1.2204597}{\bibinfo {journal} {The Journal of Chemical
  Physics}}\ }%
  \textbf{\bibinfo {volume} {124}},\ \bibinfo {pages} {219906} (\bibinfo
  {month} {Jun}\ \bibinfo {year} {2006}),\
  \url{https://doi.org/10.1063/1.2204597}%
  \bibAnnoteFile{NoStop}{JChemPhys124-219906}%
\bibitem{JChemPhys125-224106}%
  \BibitemOpen
  \bibfield{author}{%
  \bibinfo {author} {\bibfnamefont{A.~V.}\ \bibnamefont{Krukau}}, \bibinfo
  {author} {\bibfnamefont{O.~A.}\ \bibnamefont{Vydrov}}, \bibinfo {author}
  {\bibfnamefont{A.~F.}\ \bibnamefont{Izmaylov}},\ and\ \bibinfo {author}
  {\bibfnamefont{G.~E.}\ \bibnamefont{Scuseria}},\ }%
  \bibfield{journal}{%
  \Doi{10.1063/1.2404663}{\bibinfo {journal} {The Journal of Chemical
  Physics}}\ }%
  \textbf{\bibinfo {volume} {125}},\ \bibinfo {pages} {224106} (\bibinfo
  {month} {Dec}\ \bibinfo {year} {2006}),\
  \url{https://doi.org/10.1063/1.2404663}%
  \bibAnnoteFile{NoStop}{JChemPhys125-224106}%
\bibitem{JChemPhys124-221101}%
  \BibitemOpen
  \bibfield{author}{%
  \bibinfo {author} {\bibfnamefont{A.~D.}\ \bibnamefont{Becke}}\ and\ \bibinfo
  {author} {\bibfnamefont{E.~R.}\ \bibnamefont{Johnson}},\ }%
  \bibfield{journal}{%
  \Doi{http://dx.doi.org/10.1063/1.2213970}{\bibinfo {journal} {The Journal of
  Chemical Physics}}\ }%
  \textbf{\bibinfo {volume} {124}},\ \bibinfo {eid} {221101} (\bibinfo {year}
  {2006}),\
  \bibAnnoteFile{NoStop}{JChemPhys124-221101}%
\bibitem{PRB83-195134}%
  \BibitemOpen
  \bibfield{author}{%
  \bibinfo {author} {\bibfnamefont{D.}~\bibnamefont{Koller}}, \bibinfo {author}
  {\bibfnamefont{F.}~\bibnamefont{Tran}},\ and\ \bibinfo {author}
  {\bibfnamefont{P.}~\bibnamefont{Blaha}},\ }%
  \bibfield{journal}{%
  \Doi{10.1103/PhysRevB.83.195134}{\bibinfo {journal} {Phys. Rev. B}}\ }%
  \textbf{\bibinfo {volume} {83}},\ \bibinfo {pages} {195134} (\bibinfo {month}
  {May}\ \bibinfo {year} {2011}),\
  \url{http://link.aps.org/doi/10.1103/PhysRevB.83.195134}%
  \bibAnnoteFile{NoStop}{PRB83-195134}%
\bibitem{PRB85-155109}%
  \BibitemOpen
  \bibfield{author}{%
  \bibinfo {author} {\bibfnamefont{D.}~\bibnamefont{Koller}}, \bibinfo {author}
  {\bibfnamefont{F.}~\bibnamefont{Tran}},\ and\ \bibinfo {author}
  {\bibfnamefont{P.}~\bibnamefont{Blaha}},\ }%
  \bibfield{journal}{%
  \Doi{10.1103/PhysRevB.85.155109}{\bibinfo {journal} {Phys. Rev. B}}\ }%
  \textbf{\bibinfo {volume} {85}},\ \bibinfo {pages} {155109} (\bibinfo {month}
  {Apr}\ \bibinfo {year} {2012}),\
  \url{http://link.aps.org/doi/10.1103/PhysRevB.85.155109}%
  \bibAnnoteFile{NoStop}{PRB85-155109}%
\bibitem{Note2}%
  \BibitemOpen
  \bibinfo {note} {This is at variance with the results by Ghalouci \protect
  \emph {et al.}\cite {CompMatSci124-62} for InSe, obtained with the GGA,
  according to which the $\beta $-InSe is lower in energy than $\varepsilon
  $-InSe by about 100~meV per unit cell.}%
  \bibAnnoteFile{Stop}{Note2}%
\bibitem{PSSB119-327}%
  \BibitemOpen
  \bibfield{author}{%
  \bibinfo {author} {\bibfnamefont{M.}~\bibnamefont{Gatulle}}, \bibinfo
  {author} {\bibfnamefont{M.}~\bibnamefont{Fischer}},\ and\ \bibinfo {author}
  {\bibfnamefont{A.}~\bibnamefont{Chevy}},\ }%
  \bibfield{journal}{%
  \Doi{10.1002/pssb.2221190137}{\bibinfo {journal} {physica status solidi
  (b)}}\ }%
  \textbf{\bibinfo {volume} {119}},\ \bibinfo {pages} {327} (\bibinfo {month}
  {Sep}\ \bibinfo {year} {1983}),\ ISSN \bibinfo {issn} {1521-3951},\
  \url{http://dx.doi.org/10.1002/pssb.2221190137}%
  \bibAnnoteFile{NoStop}{PSSB119-327}%
\bibitem{JPCM11-6661}%
  \BibitemOpen
  \bibfield{author}{%
  \bibinfo {author} {\bibfnamefont{V.}~\bibnamefont{Panella}}, \bibinfo
  {author} {\bibfnamefont{G.}~\bibnamefont{Carlotti}}, \bibinfo {author}
  {\bibfnamefont{G.}~\bibnamefont{Socino}}, \bibinfo {author}
  {\bibfnamefont{L.}~\bibnamefont{Giovannini}}, \bibinfo {author}
  {\bibfnamefont{M.}~\bibnamefont{Eddrief}},\ and\ \bibinfo {author}
  {\bibfnamefont{C.}~\bibnamefont{S{\'e}benne}},\ }%
  \bibfield{journal}{%
  \bibinfo {journal} {Journal of Physics: Condensed Matter}\ }%
  \textbf{\bibinfo {volume} {11}},\ \bibinfo {pages} {6661} (\bibinfo {month}
  {Sep}\ \bibinfo {year} {1999}),\
  \url{http://stacks.iop.org/0953-8984/11/i=35/a=304}%
  \bibAnnoteFile{NoStop}{JPCM11-6661}%
\bibitem{Note3}%
  \BibitemOpen
  \bibinfo {note} {$B_{\unhbox \voidb@x \hbox {\relax \protect \fontsize
  {6}{7pt}\protect \selectfont [Reuss]}}= [(C_{11}\protect \tmspace
  -\thinmuskip {.1667em}+\protect \tmspace -\thinmuskip
  {.1667em}C_{12})C_{33}-2C_{13}^2]/[C_{11}\protect \tmspace -\thinmuskip
  {.1667em}+\protect \tmspace -\thinmuskip {.1667em}C_{12}\protect \tmspace
  -\thinmuskip {.1667em}+\protect \tmspace -\thinmuskip
  {.1667em}2C_{33}-4C_{13}]$; $B_{\unhbox \voidb@x \hbox {\relax \protect
  \fontsize {6}{7pt}\protect \selectfont [Voigt]}}=[2(C_{11}\protect \tmspace
  -\thinmuskip {.1667em}+\protect \tmspace -\thinmuskip
  {.1667em}C_{12})\protect \tmspace -\thinmuskip {.1667em}+\protect \tmspace
  -\thinmuskip {.1667em}4C_{13}\protect \tmspace -\thinmuskip
  {.1667em}+\protect \tmspace -\thinmuskip {.1667em}C_{33}]/9$.}%
  \bibAnnoteFile{Stop}{Note3}%
\bibitem{Note4}%
  \BibitemOpen
  \bibinfo {note} {An extension of Ref.~\protect \rev@citealpnum {PRB84-085314}
  onto 2, 3 and 4 GaSe double layers\cite {JNanoOpto7-65} provides a
  didactically nice example of bands' multiplication in a two-dimensional band
  structure.}%
  \bibAnnoteFile{Stop}{Note4}%
\bibitem{Note5}%
  \BibitemOpen
  \bibinfo {note} {The explicit superposition of three ``partial'' band
  structures, calculated with PBEsol for $\gamma $-GaSe, can be found in
  Fig.~4.4 of Ref.~\protect \rev@citealpnum {Juliana_thesis}.}%
  \bibAnnoteFile{Stop}{Note5}%
\bibitem{PSSB31-129}%
  \BibitemOpen
  \bibfield{author}{%
  \bibinfo {author} {\bibfnamefont{E.}~\bibnamefont{Aulich}}, \bibinfo {author}
  {\bibfnamefont{J.~L.}\ \bibnamefont{Brebner}},\ and\ \bibinfo {author}
  {\bibfnamefont{E.}~\bibnamefont{Mooser}},\ }%
  \bibfield{journal}{%
  \Doi{10.1002/pssb.19690310115}{\bibinfo {journal} {physica status solidi
  (b)}}\ }%
  \textbf{\bibinfo {volume} {31}},\ \bibinfo {pages} {129} (\bibinfo {year}
  {1969}),\ ISSN \bibinfo {issn} {1521-3951},\
  \url{http://dx.doi.org/10.1002/pssb.19690310115}%
  \bibAnnoteFile{NoStop}{PSSB31-129}%
\bibitem{PRB40-3837}%
  \BibitemOpen
  \bibfield{author}{%
  \bibinfo {author} {\bibfnamefont{M.}~\bibnamefont{Gauthier}}, \bibinfo
  {author} {\bibfnamefont{A.}~\bibnamefont{Polian}}, \bibinfo {author}
  {\bibfnamefont{J.~M.}\ \bibnamefont{Besson}},\ and\ \bibinfo {author}
  {\bibfnamefont{A.}~\bibnamefont{Chevy}},\ }%
  \bibfield{journal}{%
  \Doi{10.1103/PhysRevB.40.3837}{\bibinfo {journal} {Phys. Rev. B}}\ }%
  \textbf{\bibinfo {volume} {40}},\ \bibinfo {pages} {3837} (\bibinfo {month}
  {Aug}\ \bibinfo {year} {1989}),\
  \url{http://link.aps.org/doi/10.1103/PhysRevB.40.3837}%
  \bibAnnoteFile{NoStop}{PRB40-3837}%
\bibitem{MaterResBull41-751}%
  \BibitemOpen
  \bibfield{author}{%
  \bibinfo {author} {\bibfnamefont{K.}~\bibnamefont{Allakhverdiev}}, \bibinfo
  {author} {\bibfnamefont{T.}~\bibnamefont{Baykara}}, \bibinfo {author}
  {\bibfnamefont{{\c{S}}.}~\bibnamefont{Ellialtio{\u{g}}lu}}, \bibinfo {author}
  {\bibfnamefont{F.}~\bibnamefont{Hashimzade}}, \bibinfo {author}
  {\bibfnamefont{D.}~\bibnamefont{Huseinova}}, \bibinfo {author}
  {\bibfnamefont{K.}~\bibnamefont{Kawamura}}, \bibinfo {author}
  {\bibfnamefont{A.~A.}\ \bibnamefont{Kaya}}, \bibinfo {author}
  {\bibfnamefont{A.~M.}\ \bibnamefont{Kulibekov~(Gulubayov)}},\ and\ \bibinfo
  {author} {\bibfnamefont{S.}~\bibnamefont{Onari}},\ }%
  \bibfield{journal}{%
  \Doi{10.1016/j.materresbull.2005.10.015}{\bibinfo {journal} {Materials
  Research Bulletin}}\ }%
  \textbf{\bibinfo {volume} {41}},\ \bibinfo {pages} {751 } (\bibinfo {month}
  {Apr}\ \bibinfo {year} {2006}),\ ISSN \bibinfo {issn} {0025-5408},\
  \bibAnnoteFile{NoStop}{MaterResBull41-751}%
\bibitem{PhysLettA55-245}%
  \BibitemOpen
  \bibfield{author}{%
  \bibinfo {author} {\bibfnamefont{R.}~\bibnamefont{Le~Toullec}}, \bibinfo
  {author} {\bibfnamefont{M.}~\bibnamefont{Balkanski}}, \bibinfo {author}
  {\bibfnamefont{J.~M.}\ \bibnamefont{Besson}},\ and\ \bibinfo {author}
  {\bibfnamefont{A.}~\bibnamefont{Kuhn}},\ }%
  \bibfield{journal}{%
  \Doi{https://doi.org/10.1016/0375-9601(75)90729-X}{\bibinfo {journal}
  {Physics Letters {A}}}\ }%
  \textbf{\bibinfo {volume} {55}},\ \bibinfo {pages} {245} (\bibinfo {month}
  {Dec}\ \bibinfo {year} {1975}),\ ISSN \bibinfo {issn} {0375-9601},\
  \bibAnnoteFile{NoStop}{PhysLettA55-245}%
\bibitem{MaterSciEngB100-263}%
  \BibitemOpen
  \bibfield{author}{%
  \bibinfo {author} {\bibfnamefont{C.~M.}\ \bibnamefont{Julien}}\ and\ \bibinfo
  {author} {\bibfnamefont{M.}~\bibnamefont{Balkanski}},\ }%
  \bibfield{journal}{%
  \Doi{http://dx.doi.org/10.1016/S0921-5107(03)00113-2}{\bibinfo {journal}
  {Materials Science and Engineering: B}}\ }%
  \textbf{\bibinfo {volume} {100}},\ \bibinfo {pages} {263 } (\bibinfo {month}
  {jul}\ \bibinfo {year} {2003}),\ ISSN \bibinfo {issn} {0921-5107},\
  \bibAnnoteFile{NoStop}{MaterSciEngB100-263}%
\bibitem{PRB89-205416}%
  \BibitemOpen
  \bibfield{author}{%
  \bibinfo {author} {\bibfnamefont{V.}~\bibnamefont{Z\'olyomi}}, \bibinfo
  {author} {\bibfnamefont{N.~D.}\ \bibnamefont{Drummond}},\ and\ \bibinfo
  {author} {\bibfnamefont{V.~I.}\ \bibnamefont{Fal'ko}},\ }%
  \bibfield{journal}{%
  \Doi{10.1103/PhysRevB.89.205416}{\bibinfo {journal} {Phys. Rev. B}}\ }%
  \textbf{\bibinfo {volume} {89}},\ \bibinfo {pages} {205416} (\bibinfo {month}
  {May}\ \bibinfo {year} {2014}),\
  \url{https://link.aps.org/doi/10.1103/PhysRevB.89.205416}%
  \bibAnnoteFile{NoStop}{PRB89-205416}%
\bibitem{PRL80-890}%
  \BibitemOpen
  \bibfield{author}{%
  \bibinfo {author} {\bibfnamefont{Y.}~\bibnamefont{Zhang}}\ and\ \bibinfo
  {author} {\bibfnamefont{W.}~\bibnamefont{Yang}},\ }%
  \bibfield{journal}{%
  \bibinfo {journal} {Phys. Rev. Lett.}\ }%
  \textbf{\bibinfo {volume} {80}},\ \bibinfo {pages} {890} (\bibinfo {month}
  {Jan}\ \bibinfo {year} {1998})%
  \bibAnnoteFile{NoStop}{PRL80-890}%
\bibitem{PRL80-891}%
  \BibitemOpen
  \bibfield{author}{%
  \bibinfo {author} {\bibfnamefont{J.~P.}\ \bibnamefont{Perdew}}, \bibinfo
  {author} {\bibfnamefont{K.}~\bibnamefont{Burke}},\ and\ \bibinfo {author}
  {\bibfnamefont{M.}~\bibnamefont{Ernzerhof}},\ }%
  \bibfield{journal}{%
  \bibinfo {journal} {Phys. Rev. Lett.}\ }%
  \textbf{\bibinfo {volume} {80}},\ \bibinfo {pages} {891} (\bibinfo {month}
  {Jan}\ \bibinfo {year} {1998})%
  \bibAnnoteFile{NoStop}{PRL80-891}%
\bibitem{PRL102-039902}%
  \BibitemOpen
  \bibfield{author}{%
  \bibinfo {author} {\bibfnamefont{J.~P.}\ \bibnamefont{Perdew}}, \bibinfo
  {author} {\bibfnamefont{A.}~\bibnamefont{Ruzsinszky}}, \bibinfo {author}
  {\bibfnamefont{G.~I.}\ \bibnamefont{Csonka}}, \bibinfo {author}
  {\bibfnamefont{O.~A.}\ \bibnamefont{Vydrov}}, \bibinfo {author}
  {\bibfnamefont{G.~E.}\ \bibnamefont{Scuseria}}, \bibinfo {author}
  {\bibfnamefont{L.~A.}\ \bibnamefont{Constantin}}, \bibinfo {author}
  {\bibfnamefont{X.}~\bibnamefont{Zhou}},\ and\ \bibinfo {author}
  {\bibfnamefont{K.}~\bibnamefont{Burke}},\ }%
  \bibfield{journal}{%
  \Doi{10.1103/PhysRevLett.102.039902}{\bibinfo {journal} {Phys. Rev. Lett.}}\
  }%
  \textbf{\bibinfo {volume} {102}},\ \bibinfo {pages} {039902} (\bibinfo
  {month} {Jan}\ \bibinfo {year} {2009}),\
  \bibAnnoteFile{NoStop}{PRL102-039902}%
\bibitem{PRL101-239701}%
  \BibitemOpen
  \bibfield{author}{%
  \bibinfo {author} {\bibfnamefont{A.~E.}\ \bibnamefont{Mattsson}}, \bibinfo
  {author} {\bibfnamefont{R.}~\bibnamefont{Armiento}},\ and\ \bibinfo {author}
  {\bibfnamefont{T.~R.}\ \bibnamefont{Mattsson}},\ }%
  \bibfield{journal}{%
  \Doi{10.1103/PhysRevLett.101.239701}{\bibinfo {journal} {Phys. Rev. Lett.}}\
  }%
  \textbf{\bibinfo {volume} {101}},\ \bibinfo {pages} {239701} (\bibinfo
  {month} {Dec}\ \bibinfo {year} {2008}),\
  \bibAnnoteFile{NoStop}{PRL101-239701}%
\bibitem{PRL101-239702}%
  \BibitemOpen
  \bibfield{author}{%
  \bibinfo {author} {\bibfnamefont{J.~P.}\ \bibnamefont{Perdew}}, \bibinfo
  {author} {\bibfnamefont{A.}~\bibnamefont{Ruzsinszky}}, \bibinfo {author}
  {\bibfnamefont{G.~I.}\ \bibnamefont{Csonka}}, \bibinfo {author}
  {\bibfnamefont{O.~A.}\ \bibnamefont{Vydrov}}, \bibinfo {author}
  {\bibfnamefont{G.~E.}\ \bibnamefont{Scuseria}}, \bibinfo {author}
  {\bibfnamefont{L.~A.}\ \bibnamefont{Constantin}}, \bibinfo {author}
  {\bibfnamefont{X.}~\bibnamefont{Zhou}},\ and\ \bibinfo {author}
  {\bibfnamefont{K.}~\bibnamefont{Burke}},\ }%
  \bibfield{journal}{%
  \Doi{10.1103/PhysRevLett.101.239702}{\bibinfo {journal} {Phys. Rev. Lett.}}\
  }%
  \textbf{\bibinfo {volume} {101}},\ \bibinfo {pages} {239702} (\bibinfo
  {month} {Dec}\ \bibinfo {year} {2008}),\
  \bibAnnoteFile{NoStop}{PRL101-239702}%
\bibitem{JNanoOpto7-65}%
  \BibitemOpen
  \bibfield{author}{%
  \bibinfo {author} {\bibfnamefont{D.~V.}\ \bibnamefont{Rybkovskiy}}, \bibinfo
  {author} {\bibfnamefont{I.~V.}\ \bibnamefont{Vorobyev}}, \bibinfo {author}
  {\bibfnamefont{A.~V.}\ \bibnamefont{Osadchy}},\ and\ \bibinfo {author}
  {\bibfnamefont{E.~D.}\ \bibnamefont{Obraztsova}},\ }%
  \bibfield{journal}{%
  \Doi{10.1166/jno.2012.1218}{\bibinfo {journal} {Journal of Nanoelectronics
  and Optoelectronics}}\ }%
  \textbf{\bibinfo {volume} {7}},\ \bibinfo {pages} {65} (\bibinfo {month}
  {Jan}\ \bibinfo {year} {2012}),\
  \url{http://dx.doi.org/10.1166/jno.2012.1218}%
  \bibAnnoteFile{NoStop}{JNanoOpto7-65}%
\end{thebibliography}


%

\end{document}